\documentclass[12pt,a4paper]{article}
\usepackage{amsmath,graphicx,subfigure}

\newcommand{\os}{\underline}

\newcommand{\mc}{\multicolumn}
\newcommand{\pr}{\rightarrow}

\newcommand{\ba}{\begin{array}}
\newcommand{\ea}{\end{array}}

\newcommand{\vart}{\vartheta}
\newcommand{\varp}{\varphi}
\newcommand{\eps}{\varepsilon}

\newcommand{\il}{\int\limits}
\newenvironment{inspring}[1]%
{\begin{list}{}{\setlength{\rightmargin}{0cm}
                \setlength{\listparindent}{0cm}
                \settowidth{\labelwidth}{\mbox{#1}}
                \setlength{\leftmargin}{1.1\labelwidth}
                \setlength{\labelsep}{.1\labelwidth}}}%
{\end{list}}

\newcommand{\bi}[1]{\begin{inspring}{#1}}
\newcommand{\ei}{\end{inspring}}

\newcommand{\dil}{\displaystyle \int\limits}

\newcommand{\dlim}{\displaystyle \lim}

\newcommand{\dsum}{\displaystyle \sum}

\newcommand{\bfr}{{\bf r}}

\newcommand{\beq}{\begin{equation}}
\newcommand{\eq}{\end{equation}}
\catcode`\@=11

\font\tenmsa=msam10 \font\sevenmsa=msam7 \font\fivemsa=msam5
\font\tenmsb=msbm10 \font\sevenmsb=msbm7 \font\fivemsb=msbm5
\newfam\msafam
\newfam\msbfam
\textfont\msafam=\tenmsa  \scriptfont\msafam=\sevenmsa
  \scriptscriptfont\msafam=\fivemsa
\textfont\msbfam=\tenmsb  \scriptfont\msbfam=\sevenmsb
  \scriptscriptfont\msbfam=\fivemsb

\def\Bbb{\ifmmode\let\next\Bbb@\else
 \def\next{\errmessage{Use \string\Bbb\space only in math mode}}\fi\next}
\def\Bbb@#1{{\Bbb@@{#1}}}
\def\Bbb@@#1{\fam\msbfam#1}
\newcommand{\dR}{{\Bbb R}}

\newcommand{\dC}{{\Bbb C}}

\title{Zernike circle polynomials and infinite integrals involving the product of Bessel functions$^{1}$}
\author{\mbox{} \\ A.J.E.M.\ Janssen \\
Eindhoven University of Technology, \\
Department EE and EURANDOM, LG 1.39, \\
P.O.\ Box 513, 5600 MB Eindhoven, \\
The Netherlands\\
e-mail: a.j.e.m.janssen@tue.nl}
\date{}

\begin{document}
\maketitle
\footnotetext[1]{A catalogue record is available from the Eindhoven University of Technology Library \\ISBN: 978-90-386-2290-3}

\mbox{} \\ \\ \\ \\
\noindent
{\bf Abstract.} \\
Several quantities related to the Zernike circle polynomials admit an expression as an infinite integral involving the product of two or three Bessel functions. In this paper these integrals are identified and evaluated explicitly for the cases of (a)~the expansion coefficients of scaled-and-shifted circle polynomials, (b)~the expansion coefficients of the correlation of two circle polynomials, (c)~the Fourier coefficients occurring in the cosine representation of the circle polynomials, (d)~the transient response of a baffled-piston acoustical radiator due to a non-uniform velocity profile on the piston.
\newpage
\noindent
\section{Introduction} \label{sec1}
\mbox{} \\[-9mm]

Zernike circle polynomials are extensively used in the characterization of circular optical imaging systems with non-uniform pupil functions \cite{ref1}--\cite{ref8}, and, more recently, in the computation of acoustical quantities arising from harmonically excited baffled-piston radiators with non-uniform velocity profiles \cite{ref9}--\cite{ref11}. The circle polynomials were introduced by Zernike \cite{ref12} in connection with his phase contrast method and the knife-edge test. Furthermore, they played a fundamental role in Nijboer's thesis \cite{ref13} on the diffraction theory of aberrations, where they were investigated in detail.

The circle polynomials are given for integer $m$, $n$ with $n-|m|$ even and non-negative by
\beq \label{e1}
Z_n^m(\rho,\vart)=R_n^{|m|}(\rho)\,e^{im\vart}~,~~~~~~0\leq\rho\leq1\,,~~0\leq\vart<2\pi~,
\eq
where the radial polynomials $R_n^{|m|}$ are given by
\beq \label{e2}
R_n^{|m|}(\rho)=\rho^{|m|}\,P_{\bar{p}}^{(0,|m|)}(2\rho^2-1)=\dsum_{s=0}^{\bar{p}}\,({-}1)^s\,
\Bigl(\!\ba{c} n-s \\ \bar{p}\ea\!\Bigr)\,\Bigl(\!\ba{c} \bar{p} \\ s \ea\!\Bigr)\,\rho^{n-2s}~,
\eq
with $P_k^{(\alpha,\beta)}$ the general Jacobi polynomial as in \cite{ref14}, Ch.~22 and $\bar{p}=\tfrac12(n-|m|)$. It is customary to refer to $n$ as the degree and to $m$ as the azimuthal order of $Z_n^m$. The circle polynomials form a complete orthogonal system of functions on the disk $0\leq\rho\leq1$, with $Z_n^m(1,\vart)=e^{im\vart}$, that is,
\beq \label{e3}
R_n^{|m|}(1)=1~,
\eq
and the orthogonality property reads explicitly
\beq \label{e4}
\dil_0^1\!\!\dil_0^{2\pi}\,Z_n^m(\rho,\vart)((Z_{n'}^{m'}(\rho,\vart))^{\ast}\,\rho\,d\rho \,d\vart=\dfrac{\pi}{n+1}\,\,\delta_{mm'}\,\delta_{nn'}
\eq
with $\delta$ Kronecker's delta. In the sequel it will be convenient to set $R_n^{|m|}=Z_n^m\equiv0$ for integer values of $m$, $n$ such that $n-|m|$ is odd or negative.

A crucial property of the circle polynomials for diffraction theory is that their Fourier transform has the particular simple form
\begin{eqnarray}\label{e5}
& \mbox{} & \dil\hspace*{-5mm}\dil_{\nu^2+\mu^2\leq1} e^{2\pi i\nu x+2\pi i\mu y}\,Z_n^m(\rho,\vart)\,d\nu\,d\mu~= \nonumber \\[3.5mm]
& & =~\dil_0^1\!\!\dil_0^{2\pi}\,e^{2\pi i\rho r\,\cos(\vart-\varp)}\,R_n^{|m|}(\rho)\, e^{im\vart}\,\rho\,d\rho\,d\vart=
2\pi i^n\,\dfrac{J_{n+1}(2\pi r)}{2\pi r}\,e^{im\varp}~,
\end{eqnarray}
where we have written $\nu+i\mu=\rho\,e^{i\vart}$ and $x+iy=r\,e^{i\varp}$. Equivalently, in terms of Hankel transforms (of order $m$), we have
\beq \label{e6}
\dil_0^1\,R_n^{|m|}(\rho)\,J_m(2\pi r\rho)\,\rho\,d\rho= ({-}1)^{\frac{n-m}{2}}~\dfrac{J_{n+1}(2\pi r)}{2\pi r}~.
\eq
This formula was given in \cite{ref12}, Eq.~(23) and \cite{ref13}, Eq.~(2.20).

By Fourier inversion in (\ref{e4}), using Hankel transforms of order $m$, it is seen that
\beq \label{e7}
R_n^{|m|}(\rho)=({-}1)^{\frac{n-|m|}{2}}~\dil_0^{\infty}\,J_{n+1}(u)\,J_{|m|}(\rho u)\,du~,~~~~~~0\leq\rho<1~.
\eq
This result, often attributed to Noll \cite{ref6}, is shown in \cite{ref12} to follow from the discontinuous Weber-Schafheitlin integral, see \cite{ref14}, 15.4.6 on p.~561. The integral on the right-hand side of Eq.~(\ref{e7}) converges uniformly in any closed set of $\rho\geq0$ not containing 1, see Appendix~A, and its value for $\rho>1$ is 0. Thus the equality in (\ref{e7}) holds pointwise and not just in an $L^2$-sense. Also see \cite{ref15}, Appendix, Sec.~A.1.1
for a discussion of the result in Eq.~(\ref{e7}).

The results in Eqs.~(\ref{e6})--(\ref{e7}) are basic to the proof of a number of results of the circle polynomials and their radial parts. In \cite{ref6}, the result in Eq.~(\ref{e7}) was used to derive expressions for the derivative of $R_n^m$ in terms of $R$-polynomials of azimuthal orders $m\pm 1$ by employing recurrence relations for Bessel functions and their derivatives. In \cite{ref15}--\cite{ref16}, the two results in Eqs.~(\ref{e6})--(\ref{e7}) were combined with recursion properties of the Bessel functions to produce the scaling formula
\begin{eqnarray} \label{e8}
& \mbox{} & R_{n'}^m(\eps\rho)=\dsum_n\,(R_{n'}^n(\eps)-R_{n'}^{n+2}(\eps))\,R_n^m(\rho)~= \nonumber \\[3mm]
& & =~\dfrac{1}{\eps}\:\dsum_n\:\dfrac{n+1}{n'+1}\,(R_{n'+1}^{n+1}(\eps)-R_{n'-1}^{n+1}(\eps))\,R_n^m(\rho)~.
\end{eqnarray}
Here $m=0,1,...\,$, $n'=m,m+2,...\,$, and the summation is over $n=m,m+2,\dots,n'$ in which we recall the convention that $R_{n'}^{n'+2}=R_{n'-1}^{n'+1}\equiv0$ for the last term in either series. Although Eq.~(\ref{e8}) is normally used for $\eps,\rho\in[0,1]$, it should be emphasized that they are valid for all complex values of $\eps$ and $\rho$ by analyticity. The result in Eq.~(\ref{e8}) is of interest to both the lithographic community and the ophthalmological community, see \cite{ref7}, \cite{ref8}, \cite{ref15}, \cite{ref16}.

In \cite{ref17}, the formula
\beq \label{e9}
{\cal R}_n^m(p,\varp)=\dfrac{2}{n+1}\,(1-p^2)^{1/2}\,U_n(p)\,e^{im\varp}\,\chi_{[0,1)}(p)
\eq
for the Radon transform of $Z_n^m(\rho,\vart)$ is used to show that
\beq \label{e10}
R_n^m(\rho)=\dfrac1N\:\dsum_{k=0}^{N-1}\,U_n\Bigl(\rho\cos\dfrac{2\pi k}{N}\Bigr)\cos\dfrac{2\pi mk}{N}~,~~~~~~0\leq\rho\leq1~.
\eq
Here $m=0,1,...$ and $N$ is any integer $>\:n+m$, and $U_n$ is the Chebyshev polynomial of degree $n$ and second kind. This formula is interesting since it gives the $R_n^m(\rho)$ for $m\leq n<N$ in the form of a discrete cosine transform, also see \cite{ref18}. The result in Eq.~(\ref{e9}) was discovered by Cormack in \cite{ref19}, but a proof can also be based on the result in Eqs.~(\ref{e6})--(\ref{e7}) and the connection between Bessel functions and Chebyshev polynomials through the Fourier transform, see \cite{ref14}, 11.4.24--25 on p.~486.

In the present paper a number of new applications of the results in Eqs.~(\ref{e6})--(\ref{e7}) are presented. A common feature of the problems we consider is that they all give rise to infinite integrals involving the product of three Bessel functions. In Section~2 we consider the problem of finding the Zernike expansion of scaled-and-shifted circle polynomials. That is, given $a\geq0$, $b\geq0$ with $a+b\leq1$, we give explicit expressions, involving Jacobi polynomials, for the coefficients $K_{nn'}^{mm'}(a,b)$ in the expansion
\beq \label{e11}
Z_n^m(a+b\rho'e^{i\vart'})=\dsum_{n',m'}\,K_{nn'}^{mm'}(a,b)\,Z_{n'}^{m'}(\rho'e^{i\vart'})~,~~~~~~
0\leq\rho'\leq1\,,~~0\leq\vart'<2\pi~,
\eq
see Fig.~1 for the relation between the radial and angular variables of the full, centralized pupil and the reduced, shifted pupil.

\begin{figure}[h]
 \begin{center}
    \includegraphics[width = 0.5\linewidth]{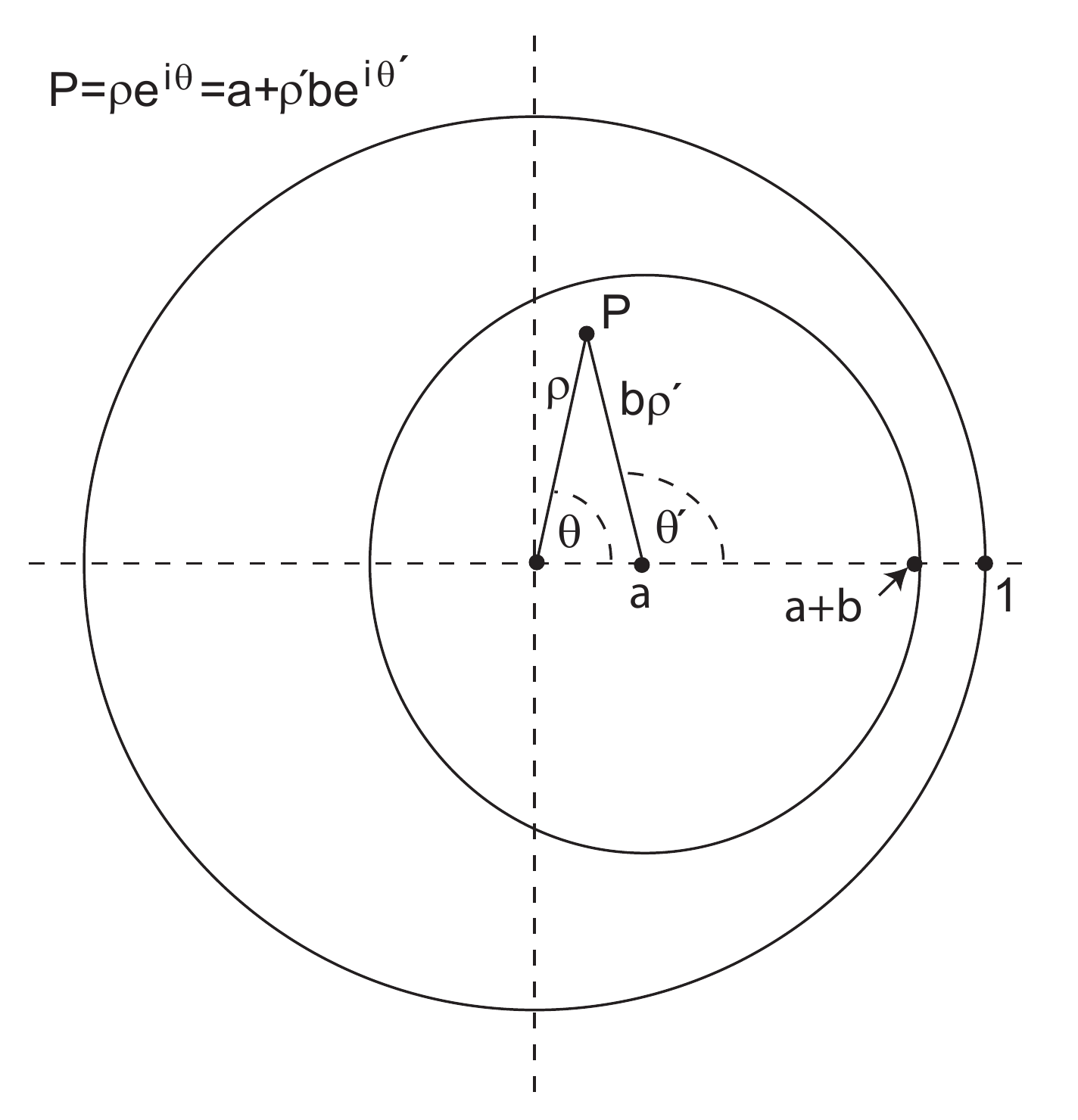}
 \end{center}
 \caption{Centralized, full pupil $\rho\,e^{i\vart}$, $0\leq\rho\leq1$, $0\leq\vart<2\pi$, and shifted and reduced pupil $a+b\,\rho'\,e^{i\vart'}$, $0\leq\rho'\leq1$, $0\leq\vart'<2\pi$, where $a\geq0$, $b\geq0$, $a+b\leq1$.}
\end{figure}

This result generalizes the scaling formulas in Eq.~(\ref{e8}) which is the case with $a=0$ in Eq.~(\ref{e11}). Furthermore, it gives the analytic solution of the transformation problem for the aberration coefficients of an eye pupil when the pupil is scaled and displaced. This problem has a long history in the ophthalmological community, see \cite{ref7} and \cite{ref20} for recent work and survey material, but no closed-form solution seems to have been found thus far. Moreover, by expressing the scaled-and-shifted polynomials as linear combinations of the orthogonal terms $Z_{n'}^{m'}$ one has a handle, via the transformation matrix elements $K_{nn'}^{mm'}(a,b)$, to tackle the important problem of assessing the condition of a finite set of circle polynomials when they are restricted to subdisks of the unit disk.

In Section~\ref{sec3} we consider the problem of computing the Zernike expansion coefficients of the correlation of two circle polynomials. Having these expansion coefficients available is of great interest when calculating transfer functions in optical imaging. The complex pupil function of the lens is expanded as a series involving the circle polynomials. The resulting series with the appropriate Zernike coefficients represent the amplitude and phase of the complex pupil function, including wavelength shift and defocusing. Through the analytic expression for the expansion coefficients of the correlation of two circle polynomials, one has direct access to the modulation transfer function over the full bandwidth of the imaging system. This mathematical device is very attractive when calculating, for instance, the pattern transfer in high-resolution optical lithography. Hence, when $m$, $n$ ,$m'$, $n'$ are integers such that $n-|m|$ and $n'-|m'|$ are even and non-negative, and when we denote for two functions $Z',Z'\in L^2(\dR^2)$ the correlation
\beq \label{e12}
(Z\,\ast\ast_{{\rm corr}}\,Z')(\nu,\mu)=\dil\!\!\dil\,Z(\nu+\nu_1,\mu+\mu_1)\,Z^{\ast}(\nu_1,\mu_1)\,d\nu_1\, d\mu_1~,
\eq
we are interested in finding the numbers $\Gamma_{nn'n''}^{mm'm''}$ such that
\beq \label{e13}
\dfrac{(n+1)(n'+1)}{\pi^2}\,(Z_n^m\,\ast\ast_{{\rm corr}}\,Z_{n'}^{m'})(\rho,\vart)=\dsum_{n'',m''}\,\dfrac
{n''+1}{4\pi}\,\Gamma_{nn'n''}^{mm'm''}\,Z_{n''}^{m''}(\tfrac12\rho,\vart)~.
\eq
Note that $Z_n^m\,\ast\ast_{{\rm corr}}\,Z_{n'}^{m'}$ is supported by the disk around 0 of radius 2 and this requires $Z_{n''}^{m''}(\tfrac12\rho,\vart)$ at the right-hand side of Eq.~(\ref{e13}). This problem has been considered in the optical context by Kintner and Sillitto in \cite{ref21}, \cite{ref22} in the interest of computing the optical transfer function (OTF) from the Zernike expansion of the pupil function. In \cite{ref21}, \cite{ref22} a number of results is obtained for the quantities $\Gamma$, but no closed-form solution is given as we obtain here. Moreover, the values of the left-hand side of Eq.~(\ref{e13}) are expressed here as integrals involving the product of three Bessel functions.

In optical simulations and/or experiments, a judicious choice of sampling points is important. It now turns out that replacing the radial variable $\rho$, $0\leq\rho\leq1$, by $\cos x$, $0\leq x\leq\pi/2$, has the effect that the variation of the radial polynomials $R_n^{|m|}$ is spread out uniformly over the $x$-range (see Fig.~6 in Section~\ref{sec7}). This suggests an adequate pupil sampling strategy. In \cite{ref23}, Subsec.~2.4, a matching procedure, using a separable set of sampling points $(\rho_k,\vart_l)$ on the disk, for estimating the Zernike expansion coefficients of a pupil from its values at the sampling points is proposed. It turns out that choosing the radial sampling points as $\cos x$, with uniformly spaced $x$ between 0 and $\pi/2$, produces near-optimal results, in the sense that the resulting method competes with Gaussian quadrature for all relevant azimuthal orders $m$ simultaneously. The fact that the variation of $R_n^{|m|}$ is spread uniformly over the $x$-range suggests, furthermore, to apply this substitution when integrals, involving the product of a radial polynomial and a function obtained from the pupil function after azimuthal integration, have to be computed. All this motivates consideration in Section~\ref{sec4} of the third problem: Finding the Fourier coefficients of the radial polynomials $R_n^{|m|}$ in their cosine-representation. Here we aim at finding the Fourier coefficients $a_{nk}^m$ in the representation
\beq \label{e14}
R_n^{|m|}(\cos x)=\dsum_{j=0}^{\lfloor n/2\rfloor}\,a_{n,n-2j}^m\cos(n-2j)\,x~.
\eq
The $a_{nk}^m$ will be found explicitly, and from this result it is seen that they are all non-negative. It then follows from $R_n^{|m|}(1)=1$, see Eq.~(\ref{e3}), that
\beq \label{e15}
|R_n^{|m|}(\rho)|\leq1~,~~~~~~0\leq\rho\leq1~.
\eq
This result was proved by Szeg\"o, see \cite{ref24}, 7.2.1 on p.~164 and the references given there, and the non-negativity of the $a_{nk}^m$ was also established by Koornwinder \cite{ref25}, Corollary~6.2 on p.~113, as was communicated to the author by Erik Koelink \cite{ref26}. The explicit form of the $a_{nk}^m$ does not seem to have been noted before. With the explicit result for the $a_{nk}^m$ available, the computation of the above mentioned integrals can be done directly and very explicitly by choosing the appropriate sampling points, of the form $\cos x$ with equidistant $x$, and using, for instance, DCT-techniques. The fact that the $a_{nk}^m$, being all non-negative with sum over $k$ equal to 1, are all small, renders this approach intrinsic stability.

The use of the Zernike circle polynomials to handle the diffraction integral for the acoustical pressure and associated quantities was proposed only recently, see \cite{ref9}--\cite{ref11}. For the case of a circular, flat radiator in an infinite baffle, the analytic results of Greenspan \cite{ref27} on on-axis pressure, far-field, power and directivity are generalized and systemized in \cite{ref9}--\cite{ref10} while \cite{ref11} deals with radiation from a flexible spherical cap on a rigid sphere. The results of Greenspan in \cite{ref26}, Section~VI, on transient responses have been briefly considered in the context of the circle polynomials in \cite{ref27}. Following Greenspan in \cite{ref26}, Secton~VI, it shall be argued in Section~\ref{sec5} that the transient response $\Phi_{\delta}(t\,;\,\bfr)$ at time $t\geq0$ and at the field point $\bfr$ due to an instantaneous volume displacement $\Delta$ at $t=0$ of the flexible piston with velocity profile $v(\sigma)$, $0\leq\sigma\leq a$ (=radius piston), is given by
\beq \label{e16}
\Phi_{\delta}(t\,;\,\bfr)=\dfrac{c\Delta H(ct-z)}{\pi\,a^2\,V_s}\,\dil_0^{\infty}\,J_0 (ub)\,J_0(uw)\,V(u)\,u\, du~.
\eq
In Eq.~(\ref{e16}), $c$ is the speed of sound, $H$ is the Heaviside function with unit step at 0, $V_s$ is the average velocity
\beq \label{e17}
V_s=\dfrac{1}{\pi a^2}\,\dil\!\!\dil_{\hspace*{-5mm}S}\,v\,dS
\eq
with $S$ the circular piston of radius $a$, and $\bfr$ is given in cylindrical coordinates as
\beq \label{e18}
\bfr=(x,y,z)=(w\cos\psi,w\sin\psi,z)~,
\eq
while
\beq \label{e19}
b=b(t\,;\,z)=\sqrt{c^2t^2-z^2}\geq0
\eq
to be considered for $ct\geq z$. Finally, $V(w)$ is the Hankel transform of order 0 of $v$, given by
\beq \label{e20}
V(u)=\dil_0^a\,J_0(u\sigma)\,v(\sigma)\,\sigma\,d\sigma~.
\eq
We restrict here to radially symmetric velocity profiles, although many of our results can be generalized to the case of non-radially symmetric profiles. When now $v(\sigma)$ is expanded as
\beq \label{e21}
v(\sigma)=V_s\,\dsum_{n=0}^{\infty}\,u_n\,R_{2n}^0(\sigma/a)~,~~~~~~0\leq\sigma\leq a~,
\eq
(Zernike expansion for radially symmetric functions on the disk using circle polynomials of azimuthal order 0), the result in Eq.~(\ref{e6}) shows that
\beq \label{e22}
\Phi_{\delta}(t\,;\,\bfr)=\dfrac{c\Delta H(ct-z)}{\pi a}\,\dsum_{n=0}^{\infty}\,({-}1)^n
\,u_n\,\dil_0^{\infty}\,J_0(ub)\,J_0(uw)\,J_{2n+1}(ua)\,du~.
\eq
Hence, again integrals of the product of three Bessel functions arise.

A word about the notation. We identify complex numbers $z$ with their polar representation $\rho\,e^{i\vart}$ or their Cartesian representation $\nu+i\mu$, whatever is most convenient in a particular setting. Thus we write things like
\beq \label{e23}
Z_n^m(a+\rho'\,b\,e^{i\vart'})~,~~~~~~Z_n^m(\nu,\mu)
\eq
to denote $Z_n^m(\rho,\vart)=R_n^{|m|}(\rho)\,e^{im\vart}$ in which
\beq \label{e24}
a+\rho'\,b\,e^{i\vart'}=\nu+i\mu=z=\rho\,e^{i\vart}
\eq
with $a,b\geq0$, $a+b\leq1$; $0\leq\rho'\leq1$, $\vart'\in[0,2\pi)$; $\nu,\mu\in\dR$, $\nu^2+\mu^2\leq1$; $0\leq\rho\leq1$, $0\leq\vart\leq 2\pi$.

\section{Scaled-and-shifted Zernike circle polynomials} \label{sec2}
\mbox{} \\[-9mm]

We shall prove the following result. \\ \\
{\bf Theorem 2.1.}~~Let $a\geq0$, $b\geq0$ with $a+b\leq1$, and let $n$, $m$ be integers with $n-|m|$ even and non-negative. Then
\beq \label{e25}
Z_n^m(a+b\,\rho'\,e^{i\vart'})=\dsum_{n',m'}\,K_{nn'}^{mm'}(a,b)\,Z_{n'}^{m'}(\rho'e^{i\vart'})~,~~~~~~
0\leq\rho'\leq1\,,~~0\leq\vart'\leq2\pi~,
\eq
where for $n=|m|,|m|+2,...\,$, $n'=|m'|,|m'|+2,...$
\beq \label{e26}
K_{nn'}^{mm'}(a,b)=T_{nn'}^{mm'}(a,b)-T_{n,n'+2}^{mm'}(a,b)~.
\eq
Here
\begin{eqnarray} \label{e27}
& \mbox{} & \hspace*{-6mm}T_{nn''}^{mm''}=({-}1)^{p-p''}\,\dil_0^{\infty}\,J_{m-m''}(au)\,J_{n''}(bu)\,J_{n+1}(u)\,du~= \nonumber \\[3mm]
& & \hspace*{-6mm}=~\left\{\!\!\ba{p{6.6cm}l}
\mc{2}{l}{\dfrac{(q{+}p'')!\,(p{-}p'')!}{(q{-}q'')!\,(p{+}q'')!}\,a^{m-m''}b^{n'}\,P_{p-p''}^{(m-m'',n'')}(1{-}2A^2)
\,P_{p-p''}^{(m-m'',n'')}(2B^2{-}1)} \\[4mm]
& \mbox{when $n-n''\geq m-m''\geq0~,$} \\[4mm]
\mc{2}{l}{\dfrac{(p{+}q'')!\,(q{-}q'')!}{(p{-}p'')!\,(q{+}p'')!}\,a^{m''-m}b^{n''}\,P_{q-q''}^{(m''-m,n'')}(1{-}2A^2)
\,P_{q-q''}^{(m''-m,n'')}(2B^2{-}1)} \\[4mm]
& \mbox{when $n-n''\geq m''-m\geq0~,$} \\[2mm]
0 & \mbox{otherwise}~.
\ea\right. \nonumber \\
& \mbox{} &
\end{eqnarray}
In Eq.~(\ref{e27}) we have written
\beq \label{e28}
p=\dfrac{n-m}{2}\,~~q=\dfrac{n+m}{2}\,,~~p''=\dfrac{n''-m''}{2}\,,~~q''=\dfrac{n''+m''}{2}~.
\eq
Furthermore, $P_k^{(\gamma,\delta)}(x)$ is the general Jacobi polynomial as in \cite{ref14}, Ch.~22 of degree $k=0,1,...$ corresponding to the weight function $(1-x)^{\gamma}(1+x)^{\delta}$ on the interval $[{-}1,1]$. Finally,
\beq \label{e29}
1-2A^2=\Bigl[(1-(a+b)^2)(1-(a-b)^2)\Bigr]^{1/2}-(a+b)(a-b)~,
\eq
\beq \label{e30}
2B^2-1={-}\Bigl[(1-(a+b)^2)(1-(a-b)^2)\Bigr]^{1/2}-(a+b)(a-b)~.
\eq
Alternatively, we have
\beq \label{e31}
A=\sin\alpha~,~~~~~~B=\sin\beta
\eq
with $\alpha\geq0$, $\beta\geq0$ such that $\alpha+\beta\leq\pi/2$ and
\beq \label{e32}
a=\sin\alpha\cos\beta~,~~~~~~b=\cos\alpha\sin\beta~.
\eq
That is, $A$ and $B$ can be obtained from the geometrical picture in Fig.~2
where $\gamma\in[\tfrac{\pi}{2},\pi]$ is such that $\sin\gamma=a+b$. \\[2mm]

\begin{figure}[h]
 \begin{center}
    \includegraphics[width = 0.6\linewidth]{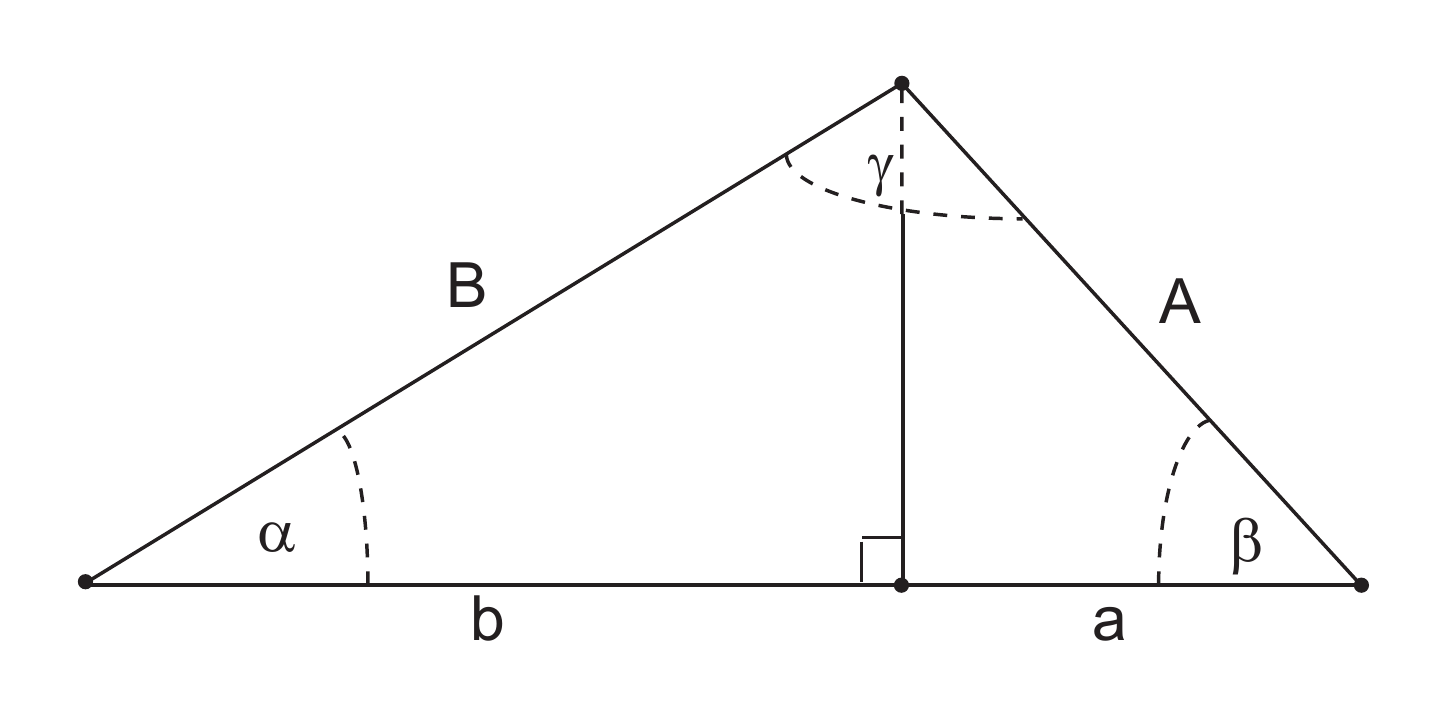}
 \end{center}
 \caption{Geometric definition of angles $\alpha$, $\beta$, $\gamma$ and side lengths $A$, $B$ from $a\geq0$, $b\geq0$, $a+b\leq1$ in accordance with Eqs.~(\ref{e31})--(\ref{e32}) and the rule of sines
$
\frac{\sin\alpha}{A}=\frac{\sin\beta}{B}=\frac{\sin\gamma}{a+b}=1~.
$}
\end{figure}

{\bf Proof.}~~By completeness and orthogonality of the circle polynomials, see Eq.~(\ref{e4}), we have that
\beq \label{e33}
K_{nn'}^{mm'}(a,b)=\dfrac{n'+1}{\pi}\,\dil_0^1\,\dil_0^{2\pi}\,Z_n^m(a+b\,\rho'\,e^{i\vart'})(Z_{n'}^{m'}
(\rho'e^{i\vart'}))^{\ast}\,\rho'\,d\rho'\,d\vart'~.
\eq
We write $\rho\,e^{i\vart}=a+\rho'\,b\,e^{i\vart'}$ in which $\rho$ and $\vart$ are depending on $\rho'$, $\vart'$ with $0\leq\rho'\leq1$, $0\leq\vart'\leq2\pi$. Then we get
\beq \label{e34}
K_{nn'}^{mm'}(a,b)=\dfrac{n'+1}{\pi}\,\dil_0^1\,\dil_0^{2\pi}\,R_n^{|m|}(\rho(\rho',\vart'))\,e^{im\vart(\rho', \vart')}\,R_{n'}^{|m'|}(\rho')\,e^{{-}m'\vart'}\,\rho'\,d\rho'\,d\vart'~.
\eq
We now use Eq.~(\ref{e7}) to rewrite $R_n^{|m|}(\rho(\rho',\vart'))$ in integral form and change the order of integration; this is allowed on account of Appendix~A where we show that the integral on the right-hand side of Eq.~(\ref{e7}) converges boundedly for all $\rho\geq0$ and uniformly, to $({-}1)^{\frac{n-|m|}{2}}\,R_n^{|m|}(\rho)$, on any set $[0,1-\eps]$ with $\eps>0$. Therefore,
\begin{eqnarray} \label{e35}
& \mbox{} & K_{nn''}^{mm'}(a,b)=\dfrac{n'+1}{\pi}\,({-}1)^{\frac{n-|m|}{2}}\:\dil_0^{\infty}\,J_{n+1}(u)\:\cdot \nonumber \\[3.5mm]
& & \cdot~\left[\dil_0^1\,\left\{\dil_0^{2\pi}\,J_{|m|}(u\rho(\rho',\vart'))\,e^{im\vart(\rho',\vart')} \,e^{-im'\vart'}\,d\vart'\right\}\cdot R_m^{|m'|}(\rho')\,\rho'\,d\rho'\right]\,du~. \nonumber \\
& \mbox{} &
\end{eqnarray}
\begin{figure}[h]
 \begin{center}
    \includegraphics[width = 0.5\linewidth]{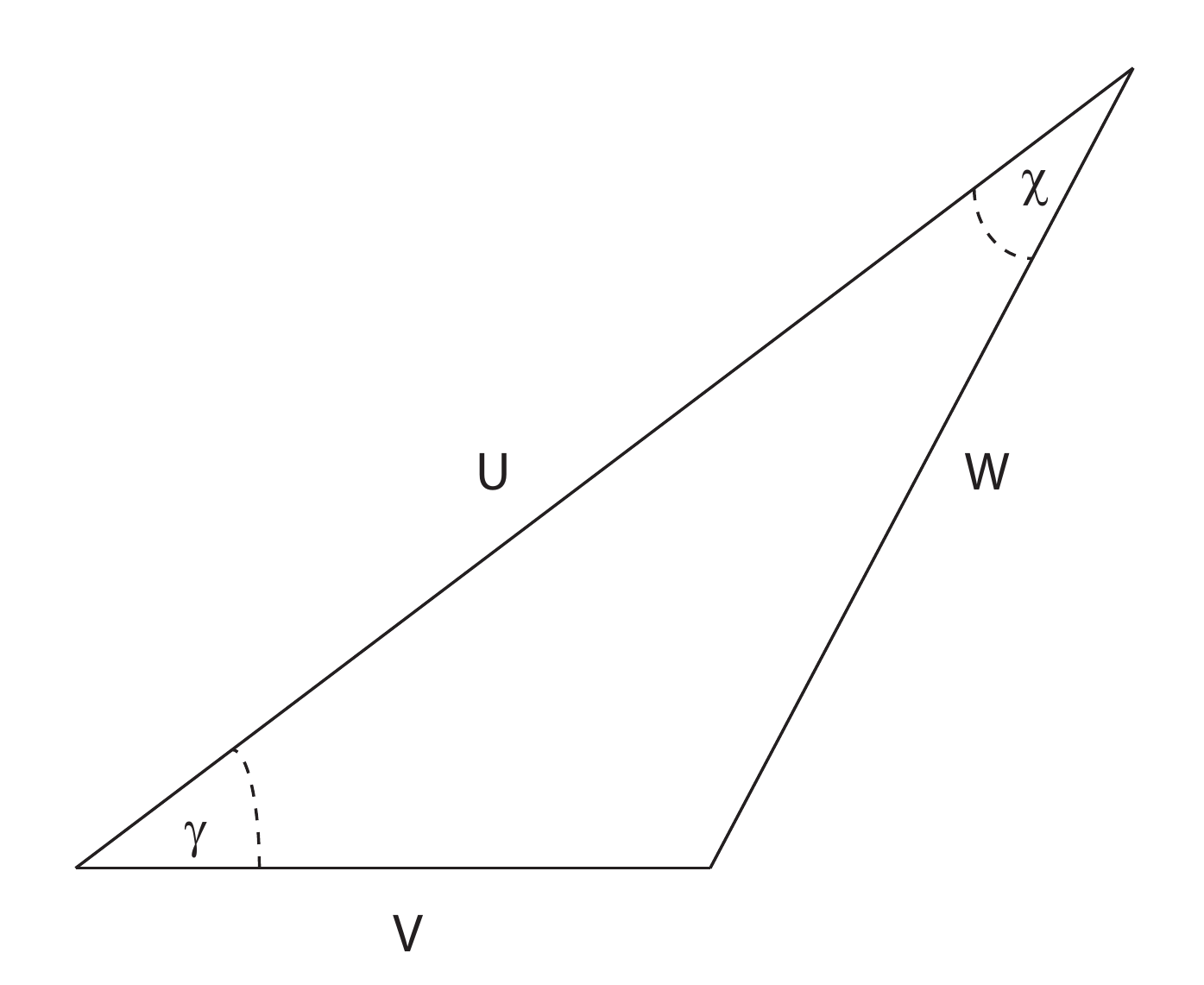}
 \end{center}
 \caption{Geometric relation between angles $\chi$, $\gamma$ and side lengths $U$, $V$, $W$ for Eq.~(\ref{e36}).}
\end{figure}

We use the addition theorem of Graf, see \cite{ref14}, 9.1.79 on p.~363,
\beq \label{e36}
{\cal C}_l(W)\,e^{il\chi}=\dsum_{k={-}\infty}^{\infty}\,{\cal C}_{l+k}(U)\,J_k(V)\,e^{ik\gamma}
\eq
for integer $l$ and ${\cal C}=J$, where $W$, $U$, $V$, $\chi$ and $\gamma$ are related as in the picture in Fig.~3.
With the variables $\rho'$, $\vart'$, $\rho$, $\vart$ as in the integral in Eq.~(\ref{e35}) and displayed in Fig.~1, we use Eq.~(\ref{e36}) with
\beq \label{e37}
W=u\,\rho(\rho',\vart')\,,~~U=ua\,,~~V=u\,\rho'\,b\,,~~\chi=\vart(\rho',\vart')\,,~~\gamma=\pi-\vart',
\eq
where we note by the comments in \cite{ref14} after 9.1.79 and 9.1.80 on p.~363 that Eq.~(\ref{e36}) can be used without any further restriction on $U$, $V$, $W$. Then we get
\beq \label{e38}
J_m(u\rho(\rho',\vart'))\,e^{im\vart(\rho',\vart')}=\dsum_{k={-}\infty}^{\infty}\,J_{m+k}(ua)\,J_k(u\,\rho'\,b)\,
e^{ik(\pi-\vart')}~.
\eq
Hence, in the case that $m\geq0$,
\begin{eqnarray} \label{e39}
& \mbox{} & \dil_0^{2\pi}\,J_m(u\rho(\rho',\vart'))\,e^{im\vart(\rho',\vart')}\,e^{-im'\vart'}\,d\vart'~= \nonumber \\[3.5mm]
& & =~2\pi({-}1)^{m'}\,J_{m-m'}(ua)\,J_{-m'}(u\,\rho'\,b)=2\pi\,J_{m-m'}(ua)\,J_{m'}(u\,\rho\,b)~, \nonumber \\
& \mbox{} &
\end{eqnarray}
and so
\begin{eqnarray} \label{e40}
& \mbox{} & \hspace*{-1cm}K_{nn'}^{mm'}(a,b)=2(n'+1)({-}1)^{\frac{n-m}{2}}\,\dil_0^{\infty}\,J_{n+1}(u)\,J_{m-m'}(ua)~\cdot \nonumber \\[3.5mm]
& & \hspace*{5cm}\cdot~\left[\dil_0^1\,J_{m'}(u\,\rho'\,b)\,R_{n'}^{|m'|}(\rho')\,d\rho'\right]\,du.
\end{eqnarray}
Using Eq.~(\ref{e38}) for $m<0$ and noting that $J_{|m|}(z)=({-}1)^m\,J_m(z)$ while
\beq \label{e41}
({-}1)^{\frac{n-|m|}{2}}\,({-}1)^m=({-}1)^{\frac{n-m}{2}}
\eq
for $m<0$, it is seen that Eq.~(\ref{e40}) holds for all integer $m$.

Next we use Eq.~(\ref{e6}) with $m'$, $n'$, $ub$ instead of $m$, $n$, $2\pi r$ to rewrite the integral in $[~]$ in Eq.~(\ref{e40}), and we obtain
\beq \label{e42}
K_{nn'}^{mm'}(a,b)=2(n'+1)({-}1)^{\frac{n+n'-m-m'}{2}}\,\dil_0^{\infty}\,\dfrac
{J_{n+1}(u)\,J_{m-m'}(ua)\,J_{n'+1}(ub)}{ub}\,du~.
\eq
Using \cite{ref14}, first item in 9.1.27 on p.~361,
\beq \label{e43}
\dfrac{J_{n'+1}(z)}{z}=\dfrac{1}{2(n'+1)}\,(J_{n'}(z)+J_{n'+2}(z))~,
\eq
it then follows that
\begin{eqnarray} \label{e44}
& \mbox{} & \hspace*{-1.5cm}K_{nn'}^{mm'}(a,b)=({-}1)^{\frac{n+n'-m-m'}{2}}\,\left[\dil_0^{\infty}\, J_{n+1}(u)\, J_{m-m'}(ua)\,J_{n'}(ub)\,du\right.~+ \nonumber \\[3.5mm]
& & \hspace*{4cm}\left.+~\dil_0^{\infty}\,J_{n+1}(u)\,J_{m-m'}(ua)\,J_{n'+2}(ub)\,du\right]~,
\end{eqnarray}
and this establishes the equality in Eq.~(\ref{e26}) with $T$'s given in integral form by the first identity in Eq.~(\ref{e27}). The second identity in Eq.~(\ref{e27}) follows from an application of a result of Bailey, the administrative details of which are deferred to Section~\ref{sec6}. \\
\mbox{}

We now list some special cases of Theorem~2.1. \\
---~~$a=0$. This gives the result of the scaling theory as developed in \cite{ref15}, \cite{ref16}, also see Eq.~(\ref{e8}). To see this, note that in Eqs.~(\ref{e31})--(\ref{e32}) we have
\beq \label{e45}
A=a=\alpha=0~,~~~~~~B=b=\sin\beta~,
\eq
and in Eq.~(\ref{e27}) only the cases with $m=m'$ give non-zero results. Furthermore,
\beq \label{e46}
P_k^{(0,n')}(1)=1~,~~~~~~b^{n'}\,P_k^{(0,n')}(2b^2-1)=R_{n'+2k}^{n'}(b)~,
\eq
and, see Eq.~(\ref{e28}),
\beq \label{e47}
p+q'=p'+q~,~~~~~~p-p'=q-q
'\eq
since $m=m'$. Plugging all this in into Eq.~(\ref{e27}) yields Eq.~(\ref{e8}). \\ \\
---~~$b=0$. Only $n'=0$ gives non-zero results in Eq.~(\ref{e27}) and then also $m'=0$ (since $|m'|\leq n'$). Now we have in Eqs.~(\ref{e31})--(\ref{e32})
\beq \label{e48}
A=a=\sin\alpha~,~~~~~~B=b=\beta=0~,
\eq
and, see Eq.~(\ref{e28}),
\beq \label{e49}
q+p'=q-q'~,~~~~~~p-p'=p+q'
\eq
since $p'=q'=m'=n'=0$. Thus, when $m\geq0$, the first case in Eq.~(\ref{e27}) yields
\beq \label{e50}
a^m\,P_p^{(m,0)}(1-2a^2)\,P_p^{(m,0)}({-}1)=a^m\,P_p^{(0,m)}(2a^2-1)=R_{m+2p}^m(a)=R_n^m(a)~,
\eq
where we have used that
\beq \label{e51}
P_k^{(\gamma,\delta)}({-}x)=({-}1)^k\,P_k^{(\delta,\gamma)}(x)~,~~~~~~P_p^{(0,m)}(1)=1~.
\eq
Therefore, we have in Eq.~(\ref{e25}) with $b=0$ the trivial representation
\beq \label{e52}
Z_n^m(a)=R_n^m(a)\,Z_0^0(\rho')~,~~~~~~0\leq\rho'\leq1\,,~~0\leq\vart'\leq2\pi~,
\eq
with a similar result in the case that $m\leq0$. \\ \\
---~~ $a+b=1$. We have in Eq.~(\ref{e31})--(\ref{e32}) in this limit case
\beq \label{e53}
\alpha+\beta=\pi/2\,,~~A=\sin\alpha\,,~~B=\cos\alpha\,,~~1-2A^2=2B^2-1=\cos 2\alpha~.
\eq
As a consequence, all $T$'s are non-negative. \\
\mbox{}

A further interesting observation is that
\beq \label{e54}
K_{nn'}^{mm'}(a,b)\neq0\Rightarrow|m'|\leq n'\leq n-|m-m'|~.
\eq
Hence, a Zernike circle polynomials $Z_n^m$ can be identified from the set of integer pairs $(m',n')$ corresponding to non-zero coefficients when $Z_n^m$ is scaled and shifted. In Subsection~7.1, a detailed computation based on Theorem~2.1 of the Zernike expansion of the scaled-and-shifted circle polynomials $Z_4^0$ and $Z_3^1$ is presented.

The validity of Theorem~2.1 will now be shown to extend to all complex values of $a$ and $b$. First assume that $a\geq0$, $b\geq0$, $a+b\leq1$. We have from Eqs.~(\ref{e1})--(\ref{e2}) that
\beq \label{e55}
Z_n^m(\rho\,e^{i\vart})=(\rho\,e^{\pm i\vart})^{|m|}\,P_{\frac{n-|m|}{2}}^{(0,|m|)}(2\rho^2-1)~,
\eq
where $\pm~=~{\rm sgn}(m)$. We have from
\begin{eqnarray} \label{e56}
\rho\,e^{\pm i\vart} & = & a+b\,\rho'\,e^{\pm i\vart'}~, \\[3mm]
\rho^2\hspace*{6mm} & = & a^2+b^2(\rho')^2+2ab\,\rho'\cos\vart'
\end{eqnarray}
that
\beq \label{e58}
Z_n^m(a+b\,\rho'\,e^{i\vart})=\dsum_{k={-}\frac{n-|m|}{2}}^{\frac{n+|m|}{2}}\,p_k(\rho'\,;\,a,b) \,e^{\pm ik\vart}~,
\eq
where $p_k(\rho'\,;\,a,b)$ depends polynomially on $\rho'$, $a$ and $b$. On the other hand, from Eqs.~(\ref{e29})--(\ref{e30}), we have that for any polynomial $p$
\beq \label{e59}
p(1-2A^2)\,p(2B^2-1)=p({-}x+y) p({-}x-y)
\eq
is an even function of $y=((1-(a+b)^2)(1-(a-b)^2))^{1/2}$ for any value of $x=(a+b)(a-b)$. Consequently, the right-hand side of Eq.~(\ref{e59}) contains only even powers of $y$. We conclude that any of the $T$'s considered in Eq.~(\ref{e27}) depends polynomially on $a$ and $b$. Hence, the relation in Eq.~(\ref{e25}) extends to all $a,b\in\dC$ by analyticity.

An important consequence of this extension is that now also the transformation matrices $(K_{nn'}^{mm'}({-}a/b,1/b))$ corresponding to the inverse transformation $z\mapsto{-}a/b+z/b$ can be considered. Accordingly, when the degrees $n$, $n'$ are restricted to a finite set $\{0,...,N\}$, the matrices corresponding to $z\mapsto a+bz$ and $z\mapsto {-}a/b+z/b$ are each other's inverse. Having expanded the shift-and-scaled circle polynomials in terms of the orthogonal functions $Z_{n'}^{m'}$, we have now the opportunity to deal with the problem of assessing the condition of the set of circle polynomials of maximal degree $N$ as a linear system when they are restricted to an arbitrary disk in the plane. Indeed, the condition number is given as the square-root of the ratio of the largest and smallest eigenvalue of the Grammian matrix, and this Grammian matrix and its inverse are expressible in terms of the appropriate transformation matrices $(K_{nn'}^{mm'}(a,b))$. Such an effort is already worthwhile for the case of restriction of circle polynomials to a disk $\rho\leq\eps$ with $\eps<1$ (pure scaling), and the author has found for this case useful and simple estimates for the magnitude of these condition numbers. This case is much simpler than the general case since the transformation matrices decouple per $m$.

\section{Zernike expansion of the optical transfer function} \label{sec3}
\mbox{} \\[-9mm]
In this section we assume that we have expanded a pupil function $P(\rho,\vart)$ (vanishing outside $\rho\leq1$) as
\beq \label{e60}
P(\rho,\vart)=\dsum_{n,m}\,\dfrac{n+1}{\pi}\,\gamma_n^m\,Z_n^m(\rho\,e^{i\vart})~,~~~~~~0\leq\rho\leq1\,,~~0\leq\vart< 2\pi~,
\eq
with coefficients $\gamma_n^m$ that can be obtained by using orthogonality of the $Z_n^m$. Writing $\nu+i\mu=\rho\,e^{i\vart}$ with $\nu,\mu\in\dR$ and identifying $P(\nu,\mu)\equiv P(\rho,\vart)$, compare end of Section~\ref{sec1}, it is required to find the Zernike expansion of the OTF (optical transfer function)
\beq \label{e61}
(P\,\ast\ast_{{\rm corr}}\,P)(\nu,\mu)=\dil\!\!\dil\,P(\nu+\nu_1,\mu+\mu_1)\,P^{\ast}(\nu_1,\mu_1)\,d\nu_1\,d\mu_1
\eq
that vanishes outside the set $\nu^2+\mu^2\leq4$, see Fig.~4. Thus, considering the expansion in Eq.~(\ref{e60}), it is required to compute (the Zernike expansion of) $Z_n^m\,\ast\ast_{{\rm corr}}\,Z_{n'}^{m'}$ for integer $n$, $m$, $n'$, $m'$ with $n-|m|$ and $n'-|m'|$ even and non-negative. We maintain the $p,q$-notation of Eq.~(\ref{e28}). \\ \\
{\bf Theorem 3.1.}~~We have, with $\nu+i\mu=\rho\,e^{i\vart}$ and $0\leq\rho\leq2$,
\begin{eqnarray} \label{e62}
& \mbox{} & (Z_n^m\,\ast\ast_{{\rm corr}}\,Z_{n'}^{m'})(\nu,\mu)~= \nonumber \\[3.5mm]
& & =2\pi({-}1)^{p-p'}\, e^{i(m-m')\vart}\, \dil_0^{\infty}\,
\dfrac{J_{n+1}(u)\,J_{n'+1}(u)\,J_{m-m'}(\rho u)}{u}\,du~,
\end{eqnarray}
and
\beq \label{e63}
\dfrac{(n+1)(n'+1)}{\pi^2}\,(Z_n^m\,\ast\ast_{{\rm corr}}\,Z_{n'}^{m'})(\rho\,e^{i\vart})= \dsum_{n'',m''}\, \dfrac{n''+1}{4\pi}\,\Gamma_{nn'n''}^{mm'm''}\,Z_{n''}^{m''}(\tfrac12\rho\,e^{i\vart})~,
\eq
where $\Gamma$ is non-vanishing for $m''=m-m'$ only, and in that case
\beq \label{e64}
\Gamma_{nn'n''}^{mm'm''}=8(n+1)(n'+1)({-}1)^{\frac{n-n'-n''}{2}}\,\dil_0^{\infty}\,J_{n+1}(u) \,J_{n'+1}(u)\, J_{n''+1}(2u)\,\dfrac{du}{u^2}~.
\eq

\begin{figure}[h]
 \begin{center}
    \includegraphics[width = 0.5\linewidth]{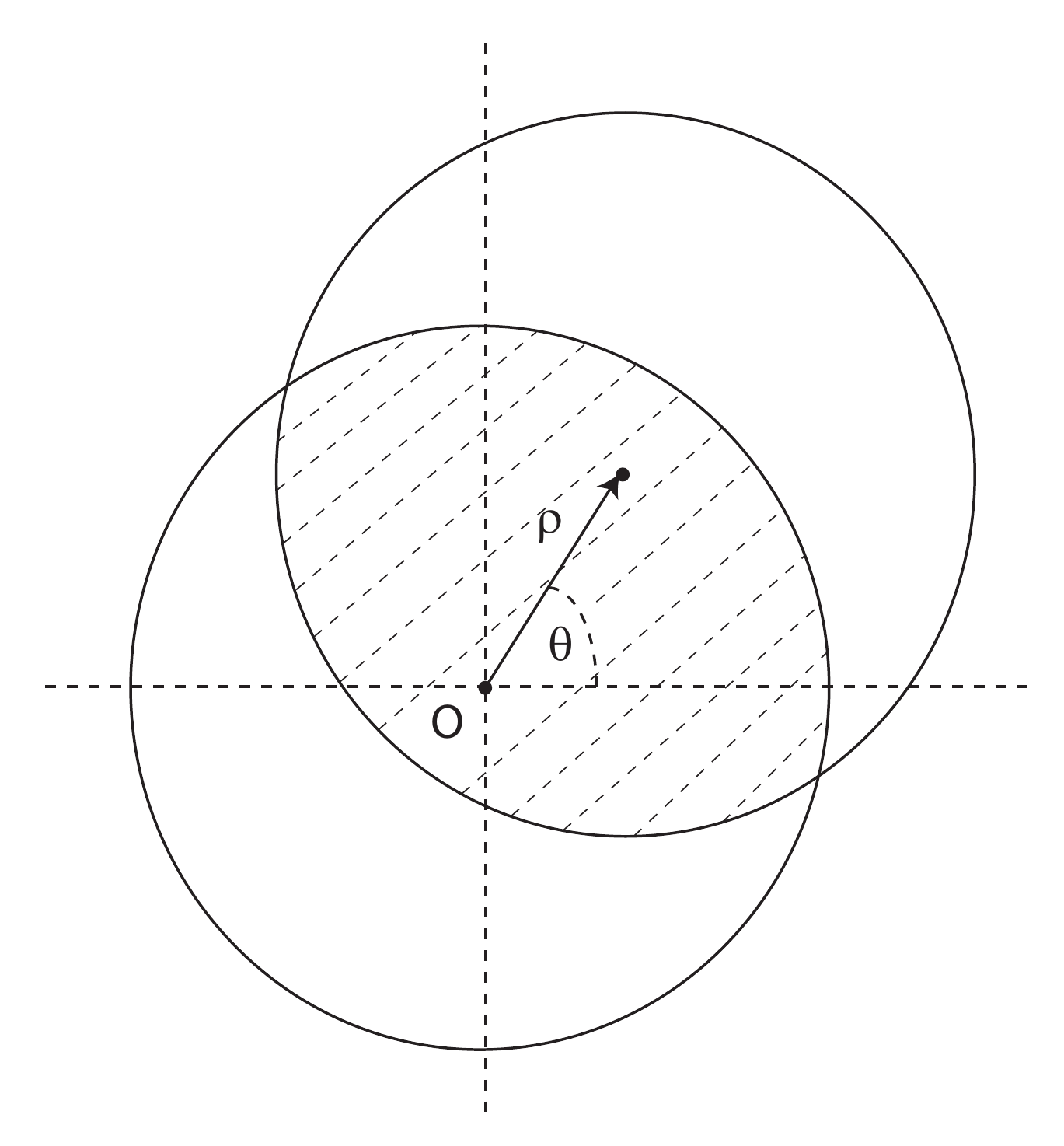}
 \end{center}
 \caption{Schematic representation of the autocorrelation function $P\,\ast\ast_{{\rm corr}}\,P$ of a non-uniform pupil function $P$ as an integral over the common region of two disks of unit radius with centers at 0 and $\rho\,e^{i\vart}$, respectively.}
\end{figure}

\noindent
{\bf Proof.}~~We have by Parseval's theorem and Eq.~(\ref{e5}) that
\begin{eqnarray} \label{e65}
& \mbox{} & (Z_n^m\,\ast\ast_{{\rm corr}}\,Z_{n'}^{m'})(\nu,\mu)~= \nonumber \\[3.5mm]
& & =~\dil\!\!\dil\,{\cal F}\,[Z_n^m(\nu+{\cdot},\mu+{\cdot})](x,y)\,({\cal F}\,Z_{n'}^{m'})^{\ast}(x,y)\,dx\,dy~= \nonumber \\[3.5mm]
& & =~\dil\!\!\dil\,e^{-2\pi i\nu x-2\pi i\mu y}({\cal F}\,Z_n^m)(x,y)\,({\cal F}\,Z_{n'}^{m'})^{\ast}(x,y)\,dx\,dy~= \nonumber \\[3.5mm]
& & =~i^{n-n'}\,\dil_0^{\infty}\dil_0^{2\pi}\, \dfrac{J_{n+1}(2\pi r)\,J_{n'+1}(2\pi r)}{r^2}\, e^{-2\pi i\rho r\cos(\vart-\varp)}\,e^{i(m-m')\varp}\,r\,dr\,d\varp~, \nonumber \\
& \mbox{} &
\end{eqnarray}
where we have written $x+iy=r\,e^{i\varp}$ and where we use that $\nu x+\mu y=\rho\,r\cos(\vart-\varp)$. Now
\beq \label{e66}
\dil_0^{2\pi}\,e^{-2\pi i\rho r\cos(\vart-\varp)}\,e^{ik\varp}\,d\varp=2\pi({-}i)^k\,e^{ik\vart}\,J_k(2\pi\rho r)~,
\eq
and inserting this into Eq.~(\ref{e65}) with $k=m-m'$, noting that $i^{n-n'}({-}i)^{m-m'}=({-}1)^{p-p'}$ and substituting $u=2\pi r$, we obtain Eq.~(\ref{e62}).

Next to show Eq.~(\ref{e64}), we note from orthogonality of the $Z_{n''}^{m''}(\tfrac12\rho\,e^{i\vart})$, see Eq.~(\ref{e4}), that
\beq \label{e67}
\Gamma_{nn'n''}^{mm'm''}=\dfrac{(n+1)(n'+1)}{\pi^2}\,\dil\!\!\dil\,(Z_n^m\,\ast\ast_{{\rm corr}}\,Z_{n'}^{m'})(\nu,\mu) (Z_{n''}^{m''}(\tfrac12\nu,\tfrac12\mu))^{\ast}\,d\nu\,d\mu~.
\eq
Again using Parseval's theorem, together with
\beq \label{e68}
{\cal F}(Z\,\ast\ast_{{\rm corr}}\,Z')={\cal F} Z\cdot({\cal F} Z')^{\ast}\,,~~({\cal F}\,Z''(\tfrac12\nu,\tfrac12\mu))(x,y)=4({\cal F} Z'')(2x,2y)~,
\eq
we obtain
\begin{eqnarray} \label{e69}
& \mbox{} & \hspace*{-7mm}\Gamma_{nn'n''}^{mm'm''}~= \nonumber \\[3mm]
& & \hspace*{-7mm}=~\dfrac{4(n+1)(n'+1)}{\pi^2}\,\dil\!\!\dil\,({\cal F}\,Z_n^m)(x,y) ({\cal F}\,Z_{n'}^{m'})^{\ast}(x,y)({\cal F}\,Z_{n''}^{m''})^{\ast}(2x,2y)\,dx\,dy~. \nonumber \\
\end{eqnarray}
Then inserting Eq.~(\ref{e5}) and using polar coordinates $x+iy=r\,e^{i\varp}$, we obtain
\begin{eqnarray} \label{e70}
& \mbox{} & \hspace*{-6.5mm}\Gamma_{nn'n''}^{mm'm''}=\dfrac{2(n+1)(n'+1)}{\pi^2}\,i^{n-n'-n''}\,\dil_0^{\infty}\,J_{n+1}(2\pi r) \,J_{n'+1}(2\pi r)\,J_{n''+1}(4\pi r)\,\dfrac{dr}{r^2}\:\cdot \nonumber \\[3.5mm]
& & \hspace*{5.5cm}\cdot~\dil_0^{2\pi}\,e^{i(m-m'-m'')\varp}\,d\varp~.
\end{eqnarray}
The proof is completed by the substitution $u=2\pi r$. Also note that $n-n'$ and $n''$ have the same parity when $m''=m-m'$. \\ \\
{\bf Notes.} \\
{\bf 1.}~~We have by Eq.~(\ref{e43}) that
\beq \label{e71}
(Z_n^m\,\ast\ast_{{\rm corr}}\,Z_{n'}^{m'})(\nu,\mu)=\dfrac{\pi}{n'+1}\,({-}1)^{p-p'}\, e^{i(m-m')\vart}\, [Q_{n+1,n'}^{m-m'}+Q_{n+1,n'+2}^{m-m'}]
\eq
and, for the case that $m''=m-m'$,
\beq \label{e72}
\Gamma_{nn'n''}^{mm'm''}=2({-}1)^{\frac{n-n'-n''}{2}}\,[Q_{nn'}^{n''+1}+Q_{n+2,n'}^{n''+1}+Q_{n,n'+2}^{n''+1} + Q_{n+2,n'+2}^{n''+1}]~,
\eq
where
\beq \label{e73}
Q_{ij}^k(a,b,c)=\dil_0^{\infty}\,J_i(au)\,J_j(bu)\,J_k(cu)\,du~,
\eq
and where the $Q$'s in Eq.~(\ref{e71}) are evaluated at $(a=1,b=1,c=\rho)$ and the $Q$'s in Eq.~(\ref{e72}) are evaluated at $(a=1,b=2,c=2)$. \\ \\
{\bf 2.}~~We shall use in Section~\ref{sec5} a result of Bailey \cite{ref28} to show the following. We have $Q_{nn'}^{n''+1}=0$ when $n''<n+n'$, and when $n''\geq n+n'$, we have
\begin{eqnarray} \label{e75}
& \mbox{} & \hspace*{-7mm}Q_{nn'}^{n''+1}(1,1,2)~= \nonumber \\[3mm]
& & \hspace*{-7mm}=~\dfrac{(\frac12(n''+n+n'))!\,(\frac12(n''-n-n'))!} {(\frac12(n''+n-n'))!\,(\frac12(n''+n'-n))!}\,(\frac12)^{n+n'+1}\, P_{\frac{n''-n-n'}{2}}^{(n,n')}(0)\, P_{\frac{n''-n-n'}{2}}^{(n',n)}(0)~, \nonumber \\
\end{eqnarray}
also see Note~2 at the end of Section~4.
\\ \\
{\bf 3.}~~From a result of Bailey \cite{ref28} it also follows that $Q_{n+1,n'}^{m-m'}$ and $Q_{n+1,n'+2}^{m-m'}$ vanish when $\rho\geq2$, but this is already clear from the fact that the $Z$'s are supported by the unit disk. The result in Eq.~(\ref{e71}) takes a more complicated form when $0<\rho<2$, see Section~\ref{sec6} for more details.\\ \\
{\bf 4.}~~In Subsection~7.2, a detailed computation, based on Theorem~3.1, for the OTF corresponding to $P = Z_0^0$ is presented.

\section{Cosine representation of the radial polynomials} \label{sec4}
\mbox{} \\[-9mm]

We shall prove the following result. \\ \\
{\bf Theorem 4.1.}~~Let $m$, $n$ be integers $\geq\:0$ with $n-m$ even and non-negative. Then
\beq \label{e76}
R_n^m(\cos x)=\dsum_{j=0}^{\lfloor n/2\rfloor}\,a_{n-2j}\cos(n-2j)\,x~,
\eq
where for integer $k\geq0$ with $n-k$ even and non-negative
\beq \label{e77}
a_k=\eps_k\,\dfrac{p!\,q!}{s!\,t!}\,(\tfrac12)^l\,(P_p^{(\gamma,\delta)}(0))^2~.
\eq
Here $\eps_0=1$, $\eps_1=\eps_2=...=2$ (Neumann's symbol), and
\beq \label{e78}
p=\dfrac{n-l}{2}\,,~~q=\dfrac{n+l}{2}\,,~~s=\dfrac{n-r}{2}\,,~~t=\dfrac{n+r}{2}\,,~~\gamma=\dfrac{l-r}{2}\,,~~
\delta=\dfrac{l+r}{2}~,
\eq
where
\beq \label{e79}
l=\max(m,k)~,~~~~~~r=\min(m,k)~.
\eq
{\bf Proof.}~~We have from Eq.~(\ref{e7})
\beq \label{e80}
R_n^m(\cos x)=({-}1)^{\frac{n-m}{2}}\,\dil_0^{\infty}\,J_{n+1}(u)\,J_m(u\cos x)\,du~.
\eq
Next we note that
\beq \label{e81}
R_n^m(\cos x)=(\cos x)^m\,P_{\frac{n-m}{2}}^{(0,m)}(\cos 2x)
\eq
has non-vanishing Fourier components $b_k\,e^{ikx}$ only for integer $k$ of the same parity as $m$. For such $k$ we shall show that
\beq \label{e82}
\dfrac{1}{2\pi}\,\dil_0^{2\pi}\,J_m(u\cos x)\,e^{ikx}\,dx=J_{\frac{m-k}{2}}(\tfrac12 u)\,J_{\frac{m+k}{2}}(\tfrac12 u)~.
\eq
Indeed, abbreviating ``the coefficient of $e^{imy}$ in'' by $C_m$, we have by the generating function
\beq \label{e83}
e^{iz\sin y}=\dsum_{m={-}\infty}^{\infty}\,J_m(z)\,e^{imy}
\eq
that
\begin{eqnarray} \label{e84}
& \mbox{} & \dfrac{1}{2\pi}\,\dil_0^{2\pi}\,J_m(u\cos x)\,e^{ikx}\,dx=\dfrac{1}{2\pi}\,C_m\,\left[ \dil_0^{2\pi}\, e^{iu\cos x\sin y}\,e^{ikx}\,dx\right]~= \nonumber \\[3.5mm]
& & =~\dfrac{1}{2\pi}\,C_m\,\left[\dil_0^{2\pi}\,e^{\frac12 iu\sin(x+y)}\,e^{-\frac12 iu\sin(x-y)}\, e^{ikx}\,dx\right]~= \nonumber \\[3.5mm]
& & =~\dfrac{1}{2\pi}\,C_m\,\left[\dsum_{m_1,m_2={-}\infty}^{\infty}\,J_{m_1}(\tfrac12 u)\, J_{m_2}(\tfrac12 u)\, \dil_0^{2\pi}\,e^{im_1(x+y)-im_2(x-y)+ikx}\,dx\right]~= \nonumber \\[3.5mm]
& & =~C_m\,\left[\dsum_{\tiny \ba{c} m_1,m_2={-}\infty, \\ m_2-m_1=k\ea}^{\infty}\,J_{m_1}(\tfrac12 u)\,J_{m_2}(\tfrac12 u)\,e^{i(m_1+m_2)y}\right]~= \nonumber \\[3.5mm]
& & =~J_{\tfrac{m-k}{2}}(\tfrac12 u)\,J_{\tfrac{m+k}{2}}(\tfrac12 u)~.
\end{eqnarray}
Then we have at once from (\ref{e80}) that
\beq \label{e85}
\dfrac{1}{2\pi}\,\dil_0^{2\pi}\,R_n^m(\cos x)\,e^{ikx}\,dx=({-}1)^{\frac{n-m}{2}}\,\dil_0^{\infty}\, J_{n+1}(u)\,J_{\frac{m-k}{2}}(\tfrac12 u)\,J_{\frac{m+k}{2}}(\tfrac12 u)\,du~.
\eq
When $k\geq0$ and $m-k\geq0$, we have then
\beq \label{e86}
\dfrac{1}{2\pi}\,\dil_0^{2\pi}\,R_n^m(\cos x)\,e^{ikx}\,dx=2({-}1)^{\frac{n-m}{2}}\,Q_{\frac{m-k}{2},\frac{m+k}{2}} ^{n+1}(1,1,2)~,
\eq
see Eqs.~(\ref{e73}) and (\ref{e75}). For this case, the result follows from Eq.~(\ref{e75}), the fact that $P_p^{(\delta,\gamma)}(0)=({-}1)^p\,P_p^{(\gamma,\delta)}(0)$ and the fact that $a_k=a_{-k}$. When $k\geq0$ and $m-k\leq0$, the result follows in a similar manner, by using that $J_{\frac{m-k}{2}}(\tfrac12 u)=({-}1)^{\frac{m-k}{2}}\,J_{\frac{k-m}{2}}(\tfrac12 u)$ and Eq.~(\ref{e75}) together with some easy administration with signs. \\ \\
{\bf Notes.} \\
{\bf 1.}~~It is straightforward to generalize the result of Theorem~4.1 to the representation of $R_n^m(v\cos x)$ with $0\leq v\leq1$. Now $Q_{\frac{m-k}{2},\frac{m+k}{2}}^{n+1}(v,v,2)$ appears in Eq.~(\ref{e86}) and $a_k$ in Eq.~(76) becomes
\beq \label{e77}
a_k(v)=\eps_k\,\dfrac{p!\,q!}{s!\,t!}\,(\tfrac12 v)^l\,P_p^{(\gamma,\delta)}(x) P_p^{(\gamma,\delta)}(-x),
\eq
where $x = (1-v^2)^{1/2}$. Also, see Section~\ref{sec6} and Eq.~(\ref{e27}). \\ \\
{\bf 2.}~~In using this result it is convenient to note that
\beq \label{e87}
P_p^{(\gamma,\delta)}(0)=\dfrac{1}{2^p}\,\dsum_{j=0}^p\,\Bigl(\!\ba{c} p+\gamma \\ j\ea\!\Bigr)\,
\Bigl(\!\ba{c} p+\delta \\ p-j\ea\!\Bigr)\,({-}1)^{p-j}~,
\eq
see \cite{ref14}, 22.3.1 on p.~775.\\ \\
{\bf 3.}~~In Subsection~7.3, a table is presented in which the radial polynomials $R_n^m$, integer $n,m$ with $0 \leq n,m \leq 8$ and $n-m \geq 0$ and even, are given in the polynomial representation of Eq.~(\ref{e2}) and in the cosine representation of Theorem~4.1.

\section{Transient response from a flexible, baffled, planar, circular piston} \label{sec5}

In this section we consider sound radiation from a circular, baffled-piston radiator with (possibly
non-uniform) velocity profile $v(\bfr_s)$ that vanishes for $|\bfr_s|>a$, where $\bfr_s$ is a point $(x_s,y_s,0)$ in the plane $z=0$ and $a$ is the radius of the piston. The complex amplitude $p(\bfr,\omega)$ of the radiated pressure at a field point $\bfr=(x,y,z)$ with $z\geq0$ due to a harmonic excitation $\exp(i\omega t)$ is given by Rayleigh's integral or King's integral (in the case that $v$ is radially symmetric) \cite{ref29}, \cite{ref30}
\begin{eqnarray} \label{e88}
p(\bfr,\omega) & = & \dfrac{ i \,\rho_0\,ck}{2\pi}\,\dil_S\,v(\bfr_s)\,\dfrac {e^{-ikr'}}{r'}\, dS~= \nonumber \\[3.5mm]
& = & i\,\rho_0\,ck\,\dil_0^{\infty}\,\dfrac{e^{-z(u^2-k^2)^{1/2}}}{(u^2-k^2)^{1/2}} \,J_0(wu)\,V(u)\,u\,du~.
\end{eqnarray}
Here $\rho_0$ is the density of the medium, $c$ is the speed of sound in the medium, $k=\omega/c$ is the wave number with $\omega$ the radian frequency, $S$ is the disk of radius $a$ centered at the origin and perpendicular to the $z$-axis, $\bfr_s$ is a point
\beq \label{e89}
(x_s,y_s,0)=(\sigma\cos\vart,\sigma\sin\vart,0)
\eq
with $0\leq\sigma\leq a$ on $S$, $\bfr$ is a field point
\beq \label{e90}
(x,y,z)=(w\cos\psi,w\sin\psi,z)
\eq
with $w\geq0$, $z\geq0$, and $r'=|\bfr-\bfr_s|$ is the distance between $\bfr$ and $\bfr_s$, see Fig.~5 for geometry and notations.
Furthermore, $(u^2-k^2)^{1/2}$ is defined to be $i\,\sqrt{k^2-u^2}$ for $0\leq u\leq k$ and $\sqrt{u^2-k^2}$ for $u\geq k$ with non-negative $\sqrt{~}$, and $V(u)$ is the Hankel transform of order 0,
\beq \label{e91}
V(u)=\dil_0^{\infty}\,J_0(u\sigma)\,v(\sigma)\,\sigma\,d\sigma~,~~~~~~u\geq0
\eq
of the radially symmetric profile $v(|\bfr_s|)$.

\begin{figure}[h]
 \begin{center}
    \includegraphics[width = 0.6\linewidth]{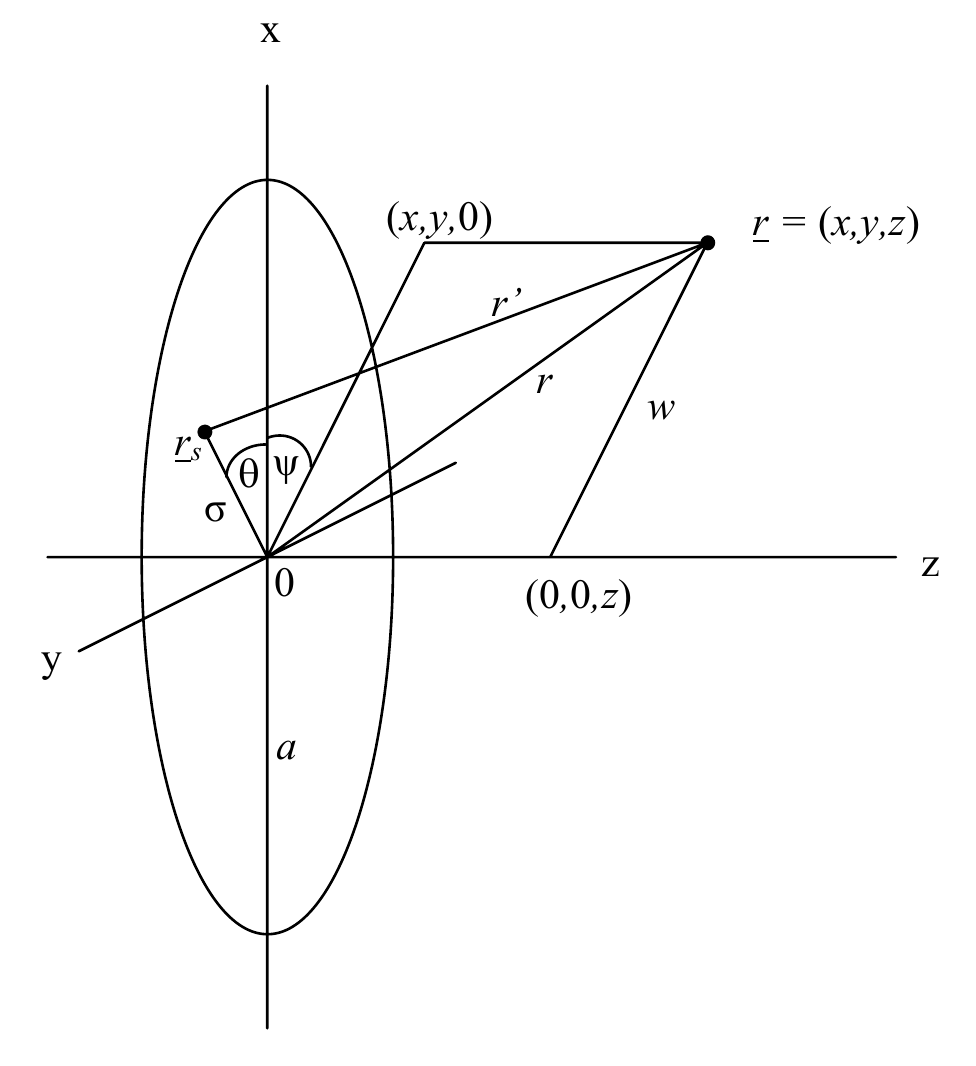}
 \caption{Geometry and notations for baffled-piston sound radiation, where $a$ is the radius of the piston and}
\begin{eqnarray}
& \mbox{} & \bfr_s=(x_s,y_s,0)=(\sigma\cos\vart,\sigma\sin\vart,0)~, \nonumber \\[3mm]
& & \bfr\hspace*{1.5mm}=(x,y,z)=(w\cos\psi,w\sin\psi,z)~, \nonumber \\[3mm]
& & r'=|\bfr-\bfr_s|=((x-x_s)^2+(y-y_s)^2+z^2)^{1/2}~. \nonumber
\end{eqnarray}
 \end{center}
\end{figure}

Bouwkamp has generalized the King integral representation to the case of non-radially symmetric profiles
\beq \label{e92}
v(\sigma,\vart)=\dsum_{m={-}\infty}^{\infty}\,v^m(\sigma)\, e^{im\vart}~,~~~~~~ 0\leq\sigma\leq a\,,~~0\leq\vart<2\pi~,
\eq
as follows. There holds \cite{ref31}
\beq \label{e93}
p(\bfr,\omega)=i\,\rho_0\,ck\,\dsum_{m={-}\infty}^{\infty}\,
\left(\dil_0^{\infty}\,\dfrac{e^{-z(u^2-k^2)^{1/2}}}{(u^2-k^2)^{1/2}}\, J_m(wu)\,V^m(u) \,u\,du\right)\,e^{im\psi}~,
\eq
where $V^m$ is the Hankel transform of order $m$ of $v^m$,
\beq \label{e94}
V^m(u)=\dil_0^{\infty}\,J_m(u\sigma)\,v^m(\sigma)\,\sigma\,d\sigma~.
\eq

In defining the transient response $\Phi_{\delta}(t\,;\,\bfr)$ at time $t\geq0$ and field point $\bfr$ due to an instantaneous volume displacement $\Delta$ at $t=0$, we follow Greenspan \cite{ref27}, also see Harris \cite{ref32}. Thus, the transient response is given by
\beq \label{e95}
\Phi_{\delta}(t\,;\,\bfr)=\dfrac{\Delta}{\pi\,a^2\,V_s}\,{\cal L}^{-1}\, \left[\Bigl. \dfrac{p(\bfr,\omega)}{i\,\rho_0\,\omega}\Bigr|_{\omega={-}is}\right]\,(t)~,
\eq
where $V_s$ denotes the average speed
\beq \label{e96}
V_s=\dfrac{1}{\pi a^2}\,\dil\!\!\dil_{\hspace*{-5mm}S}\,v(\bfr_s)\,dS~,
\eq
and ${\cal L}^{-1}$ is the inverse Laplace transform that is applied in Eq.~(\ref{e95}) to the function $p(\bfr,\omega)/i\,\rho_0\,\omega$ in which the wave number $k=\omega/c$ is replaced by $s/ic$ with $s$ the Laplace variable. Greenspan in \cite{ref27}, Section~VI uses the King integral (for the case of radially symmetric $v$), together with an identity for the inverse Laplace transform of $m^{-1}\,\exp({-}zm)$, $m=(u^2+s^2/c^2)^{1/2}$, to obtain
\beq \label{e97}
\Phi_{\delta}(t\,;\,\bfr)=\dfrac{c\,\Delta\,H(ct-z)}{\pi\,a^2\,V_s}\,\dil_0^{\infty}\,J_0 (u\,\sigma(t\,;\,z))\, J_0(uw)\,V(u)\,u\,du~,
\eq
where
\beq \label{e98}
\sigma=\sigma(t\,;\,z)=\sqrt{c^2t^2-z^2}~,~~~~~~ct\geq z~,
\eq
and $H(x)=0,1/2$ and 1 according as $x<0$, $x=0$ and $x>0$ (Heaviside function). Using Bouwkamp's generalization of the King integral representation of the pressure, see Eq.~(\ref{e93}), this yields for the transient response in Eq.~(\ref{e95})
\beq \label{e99}
\dfrac{c\,\Delta\,H(ct-z)}{\pi\,a^2\,V_s}\,\dsum_{m={-}\infty}^{\infty}\,\dil_0^{\infty}\,J_0(u\,\sigma(t\,;\,z)) \,J_m(uw)\,V^m(u)\,u\,du\cdot e^{im\psi}~.
\eq

We now develop the general velocity profile $v(\bfr_s)\equiv v(\sigma,\vart)$ into circle polynomials according to
\beq \label{e100}
v(\sigma,\vart)=V_s\,\dsum_{n,m}\,u_n^m\,R_n^{|m|}(\sigma/a)\,e^{im\vart}~,
\eq
so that $v^m(\sigma)$ is given by
\beq \label{e101}
v^m(\sigma)=V_s\,\dsum_n\,u_n^m\,R_n^{|m|}(\sigma/a)~,~~~~~~0\leq\sigma\leq a~.
\eq
Noting Eq.~(\ref{e6}), so that
\beq \label{e102}
\dil_0^a\,R_n^{|m|}(\sigma/a)\,J_m(u\sigma)\,\sigma\,d\sigma=({-}1)^{\frac{n-m}{2}}\,\dfrac{a}{u}\,J_{n+1}(ua)~,
\eq
we thus get that
\beq \label{e103}
\Phi_{\delta}(t\,;\,\bfr)=\dfrac{c\,\Delta\,H(ct-z)}{\pi a}\,\sum_{n,m}\,({-}1)^{\frac{n-m}{2}}\,u_n^m\, Q_{n+1,m}^0 (\sigma,a,w)\,e^{im\psi}~,
\eq
where
\beq \label{e104}
Q_{n+1,m}^0(\sigma,a,w)=\dil_0^{\infty}\,J_0(u\sigma)\,J_m(uw)\,J_{n+1}(ua)\,du~,
\eq
with $Q$ as in Eq.~(\ref{e73}). We shall consider the integrals at the right-hand side of Eq.~(\ref{e104}) further in Section~\ref{sec6}. Until now, except in Eq.~(\ref{e62}), the parameters $a$, $b$, $c$ occurring at the right-hand side of Eq.~(\ref{e73}) were such that they do not occur as the sides of a triangle. In that case it follows from Bailey's result in Eq.~(\ref{e110}) that $Q_{n+1,m}^0(\sigma,a,w) = 0$ with integer $n$ and $m$ such that $n-|m|$ is non-negative and even. In~(\ref{e104}) we must now allow also $\sigma,a,w > 0$ that do occur as sides of a triangle.

There are some special results for the case of radially symmetric profiles $v$. Greenspan in \cite{ref27}, Section~VI, has shown that
\beq \label{e105}
\Phi_{\delta}(t\,;\,\bfr)=\dfrac{c\,\Delta\,H(ct-z)}{\pi^2\,a^2\,V_s}\,\dil_0^A\,v\, [(w^2+\sigma^2-2w\sigma\cos\alpha)^{1/2}]\,d\alpha~,
\eq
where
\begin{eqnarray} \label{e106}
A & = & \left\{\ba{lll}
0 & \!\!, & ~~~a<|w-\sigma|~, \\[3mm]
\arccos\Bigl(\dfrac{w^2+\sigma^2-a^2}{2w\sigma}\Bigr) & \!\!, & ~~~|w-\sigma|<a<|w+\sigma|~, \\[3mm]
\pi & \!\!, & ~~~a>|w+\sigma|~,\ea\right. \nonumber \\
\end{eqnarray}
holds for velocity profiles $v^{(n)}(\sigma)$ that are a multiple of $(1-(\sigma/a)^2)^n\,H(a-\sigma)$, with $H$ the Heaviside function as before. In \cite{ref33} it has been noted that Eq.~(\ref{e105}) holds for general radially symmetric profiles $v$. Thus with
\beq \label{e107}
v(\sigma)=V_s\,\dsum_{n=0}^{\infty}\,u_{2n}^0\,R_{2n}^0(\sigma/a)~,~~~~~~0\leq\sigma\leq a~,
\eq
there follows
\beq \label{e108}
\Phi_{\delta}(t\,;\,\bfr)=\dfrac{c\,\Delta\,H(ct-z)}{\pi a}\,\dsum_{n=0}^{\infty}\, ({-}1)^n\,u_{2n}^0\, Q_{2n+1,0}^0(\sigma,a,w)~,
\eq
where, see Eqs.~(\ref{e103})--(\ref{e104}),
\begin{eqnarray} \label{e109}
({-}1)^n\,Q_{2n+1,0}^0(\sigma,a,w) & = & \dfrac{1}{\pi a}\,\dil_0^A\,R_{2n}^0 [(w^2+\sigma^2-2w\sigma\cos\alpha)^{1/2}/a]\,d\alpha~= \nonumber \\[3.5mm]
& = & \dfrac{1}{\pi a}\,\dil_0^A\,P_n\Bigl(2\,\dfrac{w^2+\sigma^2}{a^2}-1-\dfrac{4w\sigma}{a}\cos \alpha\Bigr)\,d\alpha~. \nonumber \\
\end{eqnarray}
The last expression in Eq.~(\ref{e109}) is further elaborated in \cite{ref33}, using the addition formula for Legendre polynomials, to yield a finite series expression in terms of Legendre functions and sinc.

\section{Infinite integrals involving the product of three Bessel functions} \label{sec6}
\mbox{} \\[-9mm]

We consider in this section the integrals
\beq \label{e110}
A_{\lambda\mu\nu}(a,b,c)=\dil_0^{\infty}\,J_{\lambda}(ax)\,J_{\mu}(bx)\,J_{\nu}(cx)\,dx~,
\eq
where $a,b,c>0$ and $\lambda$, $\mu$, $\nu$ are non-negative integers. There is quite some literature on these integrals and more general instances of them, see \cite{ref34}, Sec.~13.46 on pp.~411--415, \cite{ref28}, \cite{ref35}, Sec.~19.3 on pp.~349--357, \cite{ref36}, Sec.~13.4.5 on pp.~331--335. These general results can become quite unmanageable; it is the point of this section that, in the special cases that we consider, often concise and manageable results appear, often in terms of the radial polynomials themselves.

For the case that $a+b<c$ (and $a>0$, $b>0$, $c>0$), the integral in Eq.~(\ref{e110}) is given by Bailey, \cite{ref28}, Eq.~(8.1) (with a minor correction in which the $c^{\mu+\nu+1}$ in the denominator at the right-hand side should be replaced by $c$), as
\begin{eqnarray} \label{e111}
& \mbox{} & \dil_0^{\infty}\,J_{\lambda}(cu\sin\alpha\cos\beta)\,J_{\mu}(cu\cos\alpha\sin\beta)\,J_{\nu}(cu) \,du~= \nonumber \\[2.5mm]
& & =~\dfrac{\Gamma(\frac12(1+\lambda+\mu+\nu))\,\sin^{\lambda}\alpha\,\cos^{\lambda}\beta\, \cos^{\mu}\alpha\,\sin^{\mu}\beta} {c\,\Gamma(\lambda+1)\,\Gamma(\mu+1)\,\Gamma(\frac12(1-\lambda-\mu+\nu))}~\cdot \nonumber \\[3.5mm]
& & \hspace*{5mm}\cdot~{}_2F_1[\tfrac12(1+\lambda+\mu-\nu),\tfrac12(1+\lambda+\mu+\nu)\,;\, \lambda+1\,;\,\sin^2\alpha]~\cdot \nonumber \\[3.5mm]
& & \hspace*{5mm}\cdot~{}_2F_1[\tfrac12(1+\lambda+\mu-\nu),\tfrac12(1+\lambda+\mu+\nu)\,;\,\mu+1\,;\, \sin^2\beta]~,
\end{eqnarray}
where $\alpha,\beta\geq0$ are such that $\alpha+\beta<\pi/2$. In Eq.~(\ref{e27}) the choice
\beq \label{e112}
c=1\,,~~a=\sin\alpha\cos\beta\,,~~b=\cos\alpha\sin\beta~;~~~~\lambda=m-m'\,,~~\mu=n'\,,~~\nu=n+1
\eq
is made. When $m-m'\geq0$ and $A=\sin\alpha$, $B=\sin\beta$, we have
\begin{eqnarray} \label{e113}
& \mbox{} & \dil_0^{\infty}\,J_{m-m'}(au)\,J_{n'}(bu)\,J_{n+1}(u)\,du~= \nonumber \\[3.5mm]
& & =~\dfrac{\Gamma(\frac12(n+n'+m-m')+1)\,a^{m-m'}\,b^{n'}} {\Gamma(m-m'+1)\,\Gamma(n'+1)\,\Gamma(\frac12(n-n'-m+m')+1)}~\cdot \nonumber \\[3.5mm]
& & \hspace*{5mm}\cdot~{}_1F_2(\tfrac12(n'-n+m-m'),\tfrac12(n+n'+m-m')+1\,;\,m-m'+1\,;\,A^2)\,{\cdot} \nonumber \\[3.5mm]
& & \hspace*{5mm}\cdot~{}_1F_2(\tfrac12(n'-n+m-m'),\tfrac12(n+n'+m-m')+1\,;\,n'+1\,;\,B^2)~. \nonumber \\
\end{eqnarray}
Recall that both $n-m$ and $n'-m'$ are even. Hence $n-n-m'+m'$ is an even integer, and when this even integer is negative the whole expression (\ref{e113}) vanishes dues to the $\Gamma(\frac12(n-n'-m+m')+1)$ in the denominator. Using \cite{ref14}, 15.4.6 on p.~561 and 22.4.1 on p.~777, we have for $j=0,1,...$ and $\gamma,\delta\geq0$
\beq \label{e114}
{}_1F_2({-}j,\gamma+1+\delta+j\,;\,\gamma+1\,;\,x)=\dfrac{j!\,\Gamma(\gamma+1)}{\Gamma(\gamma+1+j)}\, P_j^{(\gamma,\delta)}(1-2x)~.
\eq
Therefore,
\begin{eqnarray} \label{e115}
& \mbox{} & {}_1F_2(\tfrac12(n'-n+m-m'),\tfrac12(n+n'+m-m')+1\,;\,m-m'+1\,;\,A^2)~= \nonumber \\[3.5mm]
& & =~{}_1F_2(\tfrac12(n'-n+m-m'),m-m'+1+n'~+ \nonumber \\[3.5mm]
& & \hspace*{1.4cm}+~\tfrac12(n-n'-m+m')\,;\,m-m'+1\,;\,A^2)~= \nonumber \\[3.5mm]
& & =~\dfrac{(\frac12(n'-n-m+m'))!\,(m-m')!}{(\frac12(n-n'+m-m'))!}\,P_{\frac12(n-n'-m+m')}^{(m-m',n')} (1-2A^2)~,
\end{eqnarray}
and similarly
\begin{eqnarray} \label{e116}
& \mbox{} & {}_1F_2(\tfrac12(n'-n+m-m'),\tfrac12(n+n'+m-m')+1\,;\,n'+1\,;\,B^2)~= \nonumber \\[3.5mm]
& & =~\dfrac{(\frac12(n-n'-m+m'))!\,(n')!}{(\frac12(n+n'-m+m'))!}\,P_{\frac12(n-n'-m+m')}^{(n',m-m')}(1-2B^2)~.
\end{eqnarray}
Using that $P_j^{(\gamma,\delta)}({-}x)=({-}1)^j\,P_j^{(\delta,\gamma)}(x)$ in Eq.~(\ref{e116}) and some further administration with $\Gamma$-functions and factorials then yields the first instance in Eq.~(\ref{e27}). In the case that $m'-m\geq0$, we use that $J_{m-m'}(z)=({-}1)^{m-m'}\,J_{m'-m}(z)$, and so we can apply the first instance in Eq.~(\ref{e27}) with $m$ and $m'$ interchanged. This requires a careful administration with $q$, $p$, $q'$, $p'$ as well as with the signs $({-}1)^{p-p'}$ in Eq.~(\ref{e27}). Doing so, the second instance in Eq.~(\ref{e27}) follows.

The result in Eq.~(\ref{e27}) has been proved now for the case that $a+b<1$. However, the case that $a+b=1$, $a>0$, $b>0$, follows by taking the limit case in Eq.~(\ref{e27}) and observing that the integral in Eq.~(\ref{e27}) converges uniformly in $(a,b)\in[\eps,1]\times[\eps,1]$ for any $\eps>0$ since $J_k(z)=O(z^{-1/2})$, $z\pr\infty$ and $J_k(z)$ is bounded in $z\geq0$.

We shall now consider the $Q$-integrals in Eqs.~(\ref{e73}) and Eqs.~(\ref{e85}), (\ref{e86}). These can be treated in very much the same way as the integrals in Eq.~(\ref{e113}) that arise from Eq.~(\ref{e111}) by making the choice as in Eq.~(\ref{e112}). Note that we have here the limit case $a=b=\tfrac12 c$. As to Eq.~(\ref{e75}), we note that $Q_{nn'}^{n''+1}(1,1,2)=\tfrac12 Q_{nn'}^{n''+1}(\tfrac12,\tfrac12,1)$. We thus need to replace $(m-m',n',n+1)$ by $(n,n',n''+1)$ and take
\beq \label{e117}
a=b=\tfrac12=A^2=B^2
\eq
in Eq.~(\ref{e113}). In particular, the $Q$-integral vanishes when $n'+n-n''<0$, and when $n'+n-n''\geq0$ the ${}_1F_2$ that arise in Eqs.~(\ref{e115}), (\ref{e116}) should be written down with the replacement just mentioned and the choice in Eq.~(\ref{e117}). This then yields Eq.~(\ref{e75}). Next, the $Q$-integral in Eq.~(\ref{e85}) can be handled in a similar fashion by replacing $(m-m',n',n+1)$ by $(\tfrac12(m-k),\tfrac12(m+k),n+1)$ with $a$, $b$, $A$, $B$ as in Eq.~(\ref{e113}).

We now consider the integral in Eq.~(\ref{e110}) for the case that $a,b,c\geq0$ while non of the numbers $a$, $b$, $c$ is larger than or equal to the sum of the other two; this occurs in Eqs.~(\ref{e62}), (\ref{e71}), (\ref{e73}) when $0<\rho<2$ and in Eq.~(\ref{e104}) where no other restriction on $\sigma$, $w$, $a$ is imposed than being non-negative. Here the result in \cite{ref34}, Eq.~(\ref{e7}) on p.~413 can be used. We have
\begin{eqnarray} \label{e118}
& \mbox{} & \dil_0^{\infty}\,J_{\lambda}(u\cos\varp\cos\Phi)\,J_{\mu}(u\sin\varp\sin\Phi)\,J_{\nu}(u\cos\vart)\,du~= \nonumber \\[3.5mm]
& & =~\dfrac{\cos^{\lambda}\varp\,\cos^{\lambda}\Phi\,\sin^{\mu}\varp\,\sin^{\mu}\Phi\,\cos^{\nu}\vart} {\Gamma(\nu+1)(\Gamma(\mu+1))^2}~\cdot \nonumber \\[3.5mm]
& & \hspace*{5mm}\cdot~\dsum_{k=0}^{\infty}\,({-}1)^k\,(\lambda+\mu+2k+1)\,\dfrac{\Gamma(\lambda+\mu+k+1)\, \Gamma(\mu+k+1)}{k!\,\Gamma(\lambda+k+1)}~\cdot \nonumber \\[3.5mm]
& & \hspace*{3.5cm}\cdot~\dfrac{\Gamma(\frac12(\lambda+\mu+\nu+1)+k)}{\Gamma(\frac12(\lambda+\mu-\nu+3)+k)}~\cdot \nonumber \\[3.5mm]
& & \hspace*{5mm}\cdot~{}_2F_1\Bigl(\dfrac{\nu-1-\lambda-\mu}{2}-k,\dfrac{\lambda+\mu+\nu-1}{2}+k+1\,;\,\nu+1 \,;\,\cos^2\vart\Bigr)\,{\cdot} \nonumber \\[3.5mm]
& & \hspace*{5mm}\cdot~{}_2F_1({-}k,\lambda+\mu+k+1\,;\,\mu+1\,;\,\sin^2\varp)~\cdot \nonumber \\[3.5mm]
& & \hspace*{5mm}\cdot~{}_2F_1({-}k,\lambda+\mu+k+1\,;\,\mu+1\,;\,\sin^2\Phi)
\end{eqnarray}
when ${\rm Re}(\lambda+\mu+\nu)>{-}1$ and $\cos\vart\neq{\pm}\cos(\Phi\pm\varp)$.

As to the integral in Eq.~(\ref{e104}), we use Eq.~(\ref{e118}) with $\lambda=0$, $\mu=m$, $\nu=n+1$ with $n$, $m$ non-negative integers with $n-m$ even and non-negative (the case that $m<0$ will be considered afterwards). We then have
\begin{eqnarray} \label{e119}
& \mbox{} & \hspace*{-5mm}I:=\dil_0^{\infty}\,J_0(u\cos\varp\cos\Phi)\,J_m(u\sin\varp\sin\Phi)\,J_{n+1}(u\cos\vart)\,du~= \nonumber \\[3.5mm]
& & \hspace*{-1mm}=~\dfrac{\sin^m\varp\,\sin^m\Phi\,\cos^{n+1}\vart}{\Gamma(n+2)\,\Gamma^2(m+1)}\, \dsum_{k=0}^{\infty}\,({-}1)^k(m+2k+1)\Bigl(\dfrac{\Gamma(m+k+1)}{\Gamma(k+1)}\Bigr)^2~\cdot \nonumber \\[3.5mm]
& & \hspace*{4cm}\cdot~\dfrac{\Gamma(\frac12(n+m)+k+1)} {\Gamma({-}\frac12(n-m)+k+1)}~\cdot \nonumber \\[3.5mm]
& & \hspace*{4cm}\cdot~{}_2F_1\Bigl(\dfrac{n-m}{2}-k,\dfrac{n+m}{2}+k+1\,;\,n+2\,;\,\cos^2\vart\Bigr)\,{\cdot} \nonumber \\[3.5mm]
& & \hspace*{0.6cm}\cdot~{}_2F_1({-}k,m+k+1;m+1;\sin^2\varp)\,{}_2F_1({-}k,m+k+1;m+1;\sin^2\Phi)\,. \nonumber \\
\end{eqnarray}
Next, we use (\ref{e114}) to write the product of the two last ${}_2F_1$ in Eq.~(\ref{e119}), multiplied by $\sin^m\varp\,\sin^m\Phi$, in terms of Jacobi polynomials as
\beq \label{e120}
\sin^m\varp\,P_k^{(m,0)}(1-2\sin^2\varp)\,\sin^m\Phi\,P_k^{(m,0)}(1-2\sin^2\Phi)~.
\eq
By letting $\alpha=\frac{\pi}{2}-\varp$, $\beta=\frac{\pi}{2}-\Phi$, this latter product can be written as, see Eq.~(\ref{e2}),
\beq \label{e121}
R_{m+2k}^m(\cos\alpha)\,R_{m+2k}^m(\cos\beta)~,
\eq
where it also has been used that $P_k^{(m,0)}(x)=({-}1)^k\,P_k^{(0,m)}(x)$. Furthermore, we note that the terms of the series in Eq.~(\ref{e119}) vanish when $k<\frac12(n-m)$. Letting $p=\frac12(n-m)$, writing $k=p+r$ with $r=0,1,...\,$, and noting that $m+2k=n+2r$, we see that
\begin{eqnarray} \label{e122}
& \mbox{} & I=\dsum_{r=0}^{\infty}\,({-}1)^{p+r}(n+2r+1)\,\dfrac{(n+r)!}{r!\,(n+1)!}\,\cos^{n+1}\vart~\cdot \nonumber \\[3mm]
& & \hspace*{5cm}\cdot~{}_2F_1 ({-}r,n+r+1\,;\,n+2\,;\,\cos^2\vart)\,{\cdot} \nonumber \\[3mm]
& & \hspace*{5cm}\cdot~R_{n+2r}^m(\cos\alpha)\,R_{n+2r}^m(\cos\beta)~.
\end{eqnarray}
We shall next show that
\begin{eqnarray} \label{e123}
& \mbox{} & ({-}1)^r(n+2r+1)\,\dfrac{(n+r)!}{r!\,(n+1)!}\,x^{n+1}\,{}_2F_1({-}r,n+r+1\,;\,n+2\,;\,x^2)~= \nonumber \\[3mm]
& & =~R_{n+2r+1}^{n+1}(x)-R_{n+2r-1}^{n+1}(x)~,
\end{eqnarray}
where we have $R_{n-1}^{n+1}\equiv0$ for the term with $r=0$ in the series in Eq.~(\ref{e122}). As a consequence, it follows that
\begin{eqnarray} \label{e124}
I & = & \dil_0^{\infty}\,J_0(u\sin\alpha\sin\beta)\,J_m(u\cos\alpha\cos\beta)\,J_{n+1}(u\cos\vart)\,du~= \nonumber \\[3.5mm]
& = & ({-}1)^p\,\dsum_{r=0}^{\infty}\,(R_{n+2r+1}^{n+1}(\cos\vart)-R_{n+2r-1}^{n+1}(\cos\vart))\, R_{n+2r}^m(\cos\alpha)\,R_{n+2r}^m(\cos\beta) \nonumber \\
\end{eqnarray}
under the condition that $\cos(\beta\pm\alpha)={\mp}\cos(\Phi\mp\varp)\neq\cos\vart$.

To show Eq.~(\ref{e123}), we note that for $r=0$ either side of Eq.~(\ref{e123}) equals $x^{n+1}$ by \cite{ref14}, 15.4.1 on p.~561. For $r=1,2,...\,$, we use
\begin{eqnarray} \label{e125}
R_{n+2r+1}^{n+1}(x) & = & x^{n+1}\,P_r^{(0,n+1)}(2x^2-1)=({-}1)^r\,x^{n+1}\,P_r^{(n+1,0)}(1-2x^2)~= \nonumber \\[3mm]
& = & ({-}1)^r\,x^{n+1}\,\dfrac{(n+r+1)!}{r!\,(n+1)!}\,{}_2F_1({-}r,n+r+2\,;\,n+2\,;\,x^2) \nonumber \\
\end{eqnarray}
and
\beq \label{e126}
R_{n+2r-1}^{n+1}(x)=({-}1)^{r-1}\,x^{n+1}\,\dfrac{(n+r)!}{(r-1)!\,(n+1)!}\,{}_2F_1({-}r+1,n+r+1\,;\,n+2\,;\, x^2)~.
\eq
Then the identity to be verified becomes
\begin{eqnarray} \label{e127}
& \mbox{} & (n+2r+1)\,{}_2F_1({-}r,n+r+1\,;\,n+2\,;\,x^2)~= \nonumber \\[3mm]
& & =~(n+r+1)\,{}_2F_1({-}r,n+r+2\,;\,n+2\,;\,x^2)~+ \nonumber \\[3mm]
& & \hspace*{5mm}+~r\,{}_2F_1({-}r+1,n+r+1\,;\,n+2\,;\,x^2)~,
\end{eqnarray}
and this is the contiguity relation \cite{ref14}, 15.2.14 on p.~558 with
\beq \label{e128}
a={-}r~,~~~~~~b=n+r+1~,~~~~~~c=n+2~,~~~~~~z=x^2~.
\eq
Hence Eq.~(\ref{e123}) has been established.

The result in Eq.~(\ref{e124}) has been shown for the case that $m\geq0$. When $m<0$ it holds as well when $m$ on the right-hand side of Eq.~(\ref{e124}) is replaced by $|m|$ while $p$ is maintained to be $\frac12(n-m)$. This follows from Eq.~(\ref{e124}), with $|m|$ instead of $m$, and the fact that $J_m(x)=({-}1)^m\,J_{|m|}(x)$.

Choosing $\lambda=0$, $\mu=n+1$, $\nu=m$ in Eq.~(\ref{e118}) yields
\begin{eqnarray} \label{e129}
& \mbox{} & \hspace*{-5mm}\dil_0^{\infty}\,J_0(u\sin\alpha\sin\beta)\,J_{n+1}(u\cos\alpha\cos\beta)\,J_m(u\cos\vart)\,du~= \nonumber \\[3.5mm]
& & \hspace*{-5mm}=\,({-}1)^p\,\dsum_{k=0}^{\infty}\,(R_{n+2k}^{|m|}(\cos\vart){-}R_{n+2k+2}^{|m|}(\cos\vart))\, R_{n+1+2k}^{n+1}(\cos\alpha)\,R_{n+1+2k}^{n+1}(\cos\beta)\,. \nonumber \\
\end{eqnarray}
The choice $\{\lambda,\mu\}=\{m,n+1\}$, $\nu=0$ in Eq.~(\ref{e118}) does not give an appealing result in terms of the radial polynomials, also see below.

An important special case of Eq.~(\ref{e124}) occurs when $\cos\vart=1$. Since $R_{n-1}^{n+1}(1)=0$, $R_{n+2r+1}^{n+1}(1)=1$, $r=0,1,...\,$, only the term with $r=0$ does not vanish, and there results
\begin{eqnarray} \label{e130}
& \mbox{} & \dil_0^{\infty}\,J_0(u\sin\alpha\sin\beta)\,J_m(u\cos\alpha\cos\beta)\,J_{n+1}(u)\,du~= \nonumber \\[3mm]
& & =~({-}1)^p\,R_n^m(\cos\alpha)\,R_n^m(\cos\beta)~.
\end{eqnarray}
This can be shown to be equivalent with the result in Eq.~(\ref{e27}), upper or middle case, with $m=m'$. Furthermore, the choice $\sin\alpha=0$ (so that $J_0(u\sin\alpha\sin\beta)=1$) can be shown to yield Eq.~(\ref{e7}).

We finally consider two more choices in Eq.~(\ref{e118}). The first choice has as a special case the result of Eq.~(\ref{e27}), and the second choice is of relevance for the computation of the $Q$-integrals that occur in Eq.~(\ref{e71}) for the evaluation of $Z_n^m\,\ast\ast_{{\rm corr}}\,Z_{n'}^{m'}$ at a particular point $\nu$, $\mu$.

Thus, with $\lambda=n$, $\mu=l$, $\nu=n'+1$ in Eq.~(\ref{e118}) where $n$, $l$, $n'$ are integers such that $n,n'\geq0$ while $n-n'$ and $l$ have the same parity we get in a similar manner as Eq.~(\ref{e124}) was proved the following result. There holds
\begin{eqnarray} \label{e131}
& \mbox{} & \dil_0^{\infty}\,J_n(u\cos\varp\cos\Phi)\,J_l(u\sin\varp\sin\Phi)\,J_{n'+1}(u\cos\vart)\,du~= \nonumber \\[3.5mm]
& & =~({-}1)^{\frac{n-n'+l}{2}}\,(\cos\varp\cos\Phi)^n\,(\sin\varp\sin\Phi)^{|l|}~\cdot \nonumber \\[3.5mm]
& & \hspace*{5mm}\cdot~\dsum_{k=\max\{0,\frac12(n-n'-|l|)\}}^{\infty}\,(R_{n+2k+|l|+1}^{n'+1}(\cos\vart) -R_{n+2k+|l|-1}^{n'+1}(\cos\vart))~\cdot \nonumber \\[3.5mm]
& & \hspace*{5mm}\cdot~\dfrac{\Bigl(\!\ba{c} n+k+|l| \\[-1mm] |l| \ea\!\Bigr)}
{\Bigl(\!\ba{c} k+|l| \\[-1mm] |l| \ea\!\Bigr)}\,P_k^{(n,|l|)}(1-2\sin^2\varp)\,P_k^{(n,|l|)}(1-2\sin^2\Phi)~.
\end{eqnarray}
For the case that $\vart=0$, only the term with $k=\frac12(n'-n-|l|)\geq0$ is non-vanishing, and this then yields the result of Eq.~(\ref{e27}) with $l=m-m'$ and $n$ and $n'$ interchanged.

Next, with $\lambda=n$, $\mu'=n'+1$, $\nu=l$ in Eq.~(\ref{e118}) where $n$, $l$, $n'$ are integers such that $n,n'\geq0$ while $n-n'$ and $l$ have the same parity and $n+n'\geq|l|$, we get the following result. There holds
\begin{eqnarray} \label{e132}
& \mbox{} & \dil_0^{\infty}\,J_n(u\cos\varp\cos\Phi)\,J_{n'+1}(u\sin\varp\sin\Phi)\,J_l(u\cos\vart)\,du~= \nonumber \\[3.5mm]
& & =~({-}1)^{\frac{n+n'-l}{2}}(\cos\varp\cos\Phi)^n\,(\sin\varp\sin\Phi)^{n'+1}~\cdot \nonumber \\[3.5mm]
& & \hspace*{5mm}\cdot~\dsum_{k=0}^{\infty}\,(R_{n+n'+2k}^{|l|}(\cos\vart)-R_{n+n'+2k+2}^{|l|}(\cos\vart))~\cdot \nonumber \\[3.5mm]
& & \hspace*{5mm}\cdot~\dfrac{\Bigl(\!\ba{c} n+n'+k+1 \\[-1mm] n \ea\!\Bigr)}
{\Bigl(\!\ba{c} n+k \\[-1mm] n \ea\!\Bigr)}\,P_k^{(n'+1,n)}(1-2\sin^2\varp)\,P_k^{(n'+1,n)}(1-2\sin^2\Phi)~. \nonumber \\
\end{eqnarray}
This vanishes in the case that $\cos\vart=1$. For the $Q$-integrals in Eq.~(\ref{e71}) this gives a series result by taking in Eq.~(\ref{e132})
\beq \label{e133}
\cos\varp=\cos\Phi=\sin\varp=\sin\Phi=1/\sqrt{2}~,
\eq
replacing $\frac12 u$ in the integral by $u$ and setting $2\cos\vart=\rho$.

\section{Examples} \label{sec7}
\mbox{} \\[-9mm]

In this section we present worked out examples of our main results.

\subsection{Examples for Section~\ref{sec2}} \label{subsec7.1}
\mbox{} \\[-9mm]

We use Theorem~2.1 for the computation of the Zernike expansion of the scaled-and-shifted circle polynomials $Z_4^0$ and $Z_3^1$.

\subsubsection{Computation for $Z_4^0$} \label{subsubsec7.1.1}
\mbox{} \\[-9mm]

We have
\beq \label{e134}
Z_4^0(a+b\,\rho'\,e^{i\vart'})=\dsum_{n',m'}\,K_{4n'}^{0m'}\,Z_{n'}^{m'}(\rho'\,e^{i\vart'})~,
\eq
where we have $K_{4n'}^{0m'}\,Z_{n'}^{m'}\not\equiv0$ only if $n'$ and $m'$ have the same parity and $|m'|\leq n'\leq 4-m'$, see Eq.~(\ref{e54}). This leaves us with the cases
\beq \label{e135}
|m'|=0\,,~~n'=0,2,4~;~~~~~~|m'|=1\,,~~n'=1,3~;~~~~~~|m'|=2\,,~~n'=2~.
\eq
Furthermore, $K_{4n'}^{0',{-}m'}=K_{4n'}^{0m'}$ and so it is sufficient to do the computations for the cases $m'=0,1,2$ in Eq.~(135). \\ \\
\os{$m'=0$. a.} $K_{40}^{00}=T_{40}^{00}-T_{42}^{00}$ by Eq.~(\ref{e26}) with $T_{40}^{00}$ and $T_{42}^{00}$ given by Eq.~(\ref{e27}), middle case, as
\begin{eqnarray} 
& \mbox{} & T_{40}^{00}[p=q=2\,;\,p'=q'=0]=P_2^{(0,0)}(1-2A^2)\,P_2^{(0,0)}(2B^2-1)~, \nonumber \\
& & \\
& & T_{42}^{00}[p=q=2\,;\,p'=q'=1]=b^2\,P_1^{(0,2)}(1-2A^2)\,P_1^{(0,2)}(2B^2-1)~. \nonumber \\
\end{eqnarray}
\mbox{} \\[-4mm]
\os{$m'=0$. b.} $K_{42}^{00}=T_{42}^{00}-T_{44}^{00}$ with $T_{42}^{00}$ given~in Eq.~(136) and $T_{44}^{00}$ given as
\beq \label{e138}
T_{44}^{00}(p=q=2\,;\,p'=q'=2]=b^4\,P_0^{(0,4)}(1-2A^2)\,P_0^{(0,4)}(2B^2-1)~.
\eq
\mbox{} \\[-4mm]
\os{$m'=0$. c.} $K_{42}^{00}=T_{44}^{00}-T_{46}^{00}$ with $T_{44}^{00}$ given~in Eq.~(\ref{e138}) and $T_{46}^{00}=0$. \\ \\
\os{$m'=1$. a.} $K_{41}^{01}=T_{41}^{01}-T_{43}^{01}$ by Eq.~(\ref{e26}) with $T_{41}^{01}$ and $T_{43}^{01}$ given by Eq.~(\ref{e27}), middle case, as
\begin{eqnarray} 
& \mbox{} & T_{41}^{01}[p=q=2\,;\,p'=0,q'=1]=\tfrac32 ab\,P_1^{(1,1)}(1-2A^2)\,P_1^{(1,1)}(2B^2-1)~, \nonumber \\[2mm]
& & \\
& & T_{43}^{01}[p=q=2\,;\,p'=1,q'=2]=4ab^3\,P_0^{(1,3)}(1-2A^2)\,P_0^{(1,3)}(2B^2-1)~. \nonumber \\
\end{eqnarray}
\mbox{} \\[-4mm]
\os{$m'=1$. b.} $K_{43}^{01}=T_{43}^{01}-T_{45}^{01}$ with $T_{43}^{01}$ given by Eq.~(139) and $T_{45}^{01}=0$. \\ \\
\os{$m'=2$. a.} $K_{42}^{02}=T_{42}^{02}-T_{44}^{02}$ by Eq.~(\ref{e26}) with $T_{42}^{02}$ given by Eq.~(\ref{e27}), middle case, as
\beq \label{e141}
T_{42}^{02}=[p=q=2\,;\,p'=0,q'=2]=6a^2b^2\,P_0^{(2,2)}(1-2A^2)\,P_0^{(2,2)}(2B^2-1)~,
\eq
and $T_{44}^{02}=0$ since $n-n''=0<2=|m-m''|$, see Eq.~(\ref{e27}). \\
\mbox{}

There remains to be calculated the right-hand side of Eqs.~(135)--(140) with $P_k^{(\gamma,\delta)}$ the Jacobi polynomials and $1-2A^2$ and $2B^2-1$ given in terms of $a$ and $b$ by Eqs.~(\ref{e29})--(\ref{e30}). In general, one can use that
\beq \label{e142}
P_k^{(\gamma,\delta)}(x)=\dfrac{(k+\gamma)!}{k!\,(k+\gamma+\delta)!}\,\dsum_{l=0}^k\,\Bigl(\!\ba{c} k \\[-1mm] l \ea\!\Bigr)\,\dfrac{(k+l+\gamma+\delta)!}{2^l(l+\gamma)!}\,(x-1)^l~,
\eq
together with $P_k^{(\gamma,\delta)}({-}x)=({-}1)^k\,P_k^{(\delta,\gamma)}(x)$. For the present purposes it is sufficient to know that
\beq \label{e143}
P^{(0,0)}(x)=\tfrac32 x^2-1~;~~~~~~P_1^{(0,\delta)}(x)=(1+\tfrac12\delta)\,x-\tfrac12\delta~;~~~~~~P_0^{(\gamma,\delta)}(x)=1~.
\eq
Using Eqs.~(\ref{e143}) and (\ref{e29}), (\ref{e30}) in Eqs.~(135)--(140) yields
\begin{eqnarray} 
& \mbox{} & T_{40}^{00}=6a^4+6b^4+24a^2b^2-6a^2-6b^2+1~, \nonumber \\[3mm]
& & T_{42}^{00}=b^2(12a^2+4b^2-3)~,~~~~~~T_{44}^{00}=b^4~, \\[3mm]
& & T_{41}^{01}=\tfrac32 ab(8a^2+8b^4-4)~,~~~~~T_{43}^{01}=4ab^3~, \\[3mm]
& & T_{42}^{02}=6a^2b^2~.
\end{eqnarray}
This then gives
\begin{eqnarray} 
& \mbox{} & K_{40}^{00}=6a^4+2b^4+12a^2b^2-6a^2-3b^2+1~, \nonumber \\[3mm]
& & K_{42}^{00}=3b^4+12a^2b^2-3b^2~,~~~~~~K_{44}^{00}=b^4~, \\[3mm]
& & K_{41}^{01}=12a^3b+8ab^3-6ab~,~~~~~K_{43}^{01}=4ab^3~, \\[3mm]
& & K_{42}^{02}=6a^2b^2~.
\end{eqnarray}
Hence
\begin{eqnarray} \label{e148}
& \mbox{} & Z_4^0(a+b\,\rho'\,e^{i\vart'})=[(6a^4+2b^4+12a^2b^2-6a^2-3b^2+1)\,Z_0^0~+ \nonumber \\[3mm]
& & +~(3b^4+12a^2b^2-3b^2)\,Z_2^0+b^4Z_4^0]+[(12a^3+8ab^3-6ab)\,Z_1^1~+ \nonumber \\[3mm]
& & +~(12a^3b+8ab^3-6ab)\,Z_1^{-1}+4ab^3\,Z_3^1+4ab^3\,Z_3^{-1}]~+ \nonumber \\[3mm]
& & +~[6a^2b^2\,Z_2^2+6a^2b^2]\,Z_2^{-2}~,
\end{eqnarray}
where the $Z_{n'}^{m'}$ at the right-hand side of Eq.~(\ref{e148}) should be evaluated at $\rho'\,e^{i\vart'}$.

\subsubsection{Computation for $Z_3^1$} \label{subsubsec7.1.2}
\mbox{} \\[-9mm]

We have
\beq \label{e149}
Z_3^1(a+b\,\rho'\,e^{i\vart'})=\dsum_{n',m'}\,K_{3n'}^{1m'}\,Z_{n'}^{m'}(\rho'\,e^{i\vart'})~,
\eq
where we have $K_{3n'}^{1m'}\,Z_{n'}^{m'}\not\equiv0$ only if $n'$ and $m'$ have the same parity and $|m'|\leq n'\leq3-|m'-1|$. This leaves us with the cases
\beq \label{e150}
m'={-}1,~n'=1~;~~~m'=0,~n'=0,2~;~~~m'=1,~n'=1,3~;~~~m'=2\,,~n'=2\,.
\eq
Thus we compute subsequently \\ \\
\os{$m'={-}1$. a.} $K_{31}^{1,{-}1}=T_{31}^{1,{-}1}-T_{33}^{1,{-}1}$ by Eq.~(\ref{e26}) with $T_{31}^{1,{-}1}$ given by Eq.~(\ref{e27}), upper case, as
\beq \label{e151}
T_{31}^{1,{-}1}[p=1,q=2\,;\,p'=1,q'=0]=3a^2b\,P_0^{(2,1)}(1-2A^2)\,P_0^{(2,1)}(2B^2-1)~,
\eq
and $T_{33}^{1,{-}1}=0$ since $n-n''=0<4=|m-m''|$, see Eq.~(\ref{e27}). \\ \\
\os{$m'=0$. a.} $K_{30}^{10}=T_{30}^{10}-T_{32}^{10}$ with $T_{30}^{10}$ and $T_{32}^{10}$ given by Eq.~(\ref{e27}), upper case, as
\begin{eqnarray} 
& \mbox{} & T_{30}^{10}[p=1,q=2\,;\,p'=0,q'=0]=a\,P_1^{(1,0)}(1-2A^2)\,P_1^{(1,0)}(2B^2-1)~, \nonumber \\
& & \\
& & T_{32}^{10}[p=1,q=2\,;\,p'=1,q'=1]=3ab^2\,P_0^{(1,2)}(1-2A^2)\,P_0^{(1,2)}(2B^2-1)~. \nonumber \\
\end{eqnarray}
\os{$m'=0$. b.} $K_{32}^{10}=T_{32}^{10}-T_{34}^{10}$ with $T_{32}^{10}$ given in Eq.~(154) and $T_{34}^{10}=0$ since $n-n''={-}1<1=|m-m''|$, see Eq.~(\ref{e27}). \\ \\
\os{$m'=1$. a.} $K_{31}^{11}=T_{31}^{11}-T_{33}^{11}$ with $T_{31}^{11}$ and $T_{33}^{11}$ given by Eq.~(\ref{e27}), both upper and middle case, as
\begin{eqnarray} 
& \mbox{} & T_{31}^{11}[p=1,q=2\,;\,p'=0,q'=1]=b\,P_1^{(0,1)}(1-2A^2)\,P_1^{(0,1)}(2B^2-1)~, \nonumber \\
& & \\
& & T_{33}^{11}[p=1,q=2\,;\,p'=1,q'=2]=b^3\,P_0^{(0,3)}(1-2A^2)\,P_0^{(0,3)}(2B^2-1)~. \nonumber \\
\end{eqnarray}
\os{$m'=1$. b.} $K_{33}^{11}=T_{33}^{11}-T_{35}^{11}$ with $T_{33}^{11}$ given by Eq.~(156) and $T_{35}^{11}=0$. \\ \\
\os{$m'=2$. a.} $K_{32}^{12}=T_{32}^{12}-T_{34}^{12}$ with $T_{32}^{12}$ given by Eq.~(\ref{e27}), middle case, as
\beq \label{e156}
T_{32}^{12}[p=1,q=2\,;\,p'=0,q'=2]=3ab^2\,P_0^{(1,2)}(1-2A^2)\,P_0^{(1,2)}(2B^2-1)~,
\eq
and $T_{34}^{12}=0$. \\ \\
We calculate the right-hand sides of Eqs.~(152)--(157) as
\begin{eqnarray} 
& \mbox{} & T_{31}^{1,{-}1}=3a^2b~, \\[3mm]
& & T_{30}^{10}\hspace*{3mm}=a(3a^2+6b^2-2)~,~~~~~~T_{32}^{10}=3ab^2~, \\[3mm]
& & T_{31}^{11}\hspace*{3mm}=b(6a^2+3b^2-2)~,~~~~~~T_{33}^{11}=b^3~, \\[3mm]
& & T_{32}^{12}\hspace*{3mm}=3ab^2~.
\end{eqnarray}
This gives
\begin{eqnarray} 
& \mbox{} & K_{31}^{1,{-}1}=3a^2b~, \nonumber \\[3mm]
& & K_{30}^{10}\hspace*{3mm}=3a^3+3ab^2-2a~,~~~~~K_{32}^{10}=3ab^2~, \\[3mm]
& & K_{31}^{11}\hspace*{3mm}=6a^2b+2b^3-2b~,~~~~~~K_{33}^{11}=b^3~, \\[3mm]
& & K_{32}^{12}\hspace*{3mm}=3ab^2~. \nonumber
\end{eqnarray}
Hence
\begin{eqnarray} \label{e163}
& \mbox{} & Z_3^1(a+b\,\rho'\,e^{i\vart'})=3a^2b\,Z_1^{-1}+[(3a^3+3ab^2-2a)\,Z_0^0+3ab^2\,Z_2^0]~+ \nonumber \\[3mm]
& & +~[(6a^2b+2b^3-2b)\,Z_1^1+b^3\,Z_3^1]+3ab^2\,Z_2^2~,
\end{eqnarray}
where the $Z_{n'}^{m'}$ at the right-hand side of Eq.~(\ref{e164}) should be evaluated at $\rho'\,e^{i\vart'}$. \\
\mbox{}

It is obvious that for the expansion of a general $Z_n^m(a+b\,\rho'\,e^{i\vart'})$ one can construct a concise and efficient computer code on basis of Theorem~2.1, taking advantage of the various shortcuts and reuse of intermediate results such as those encountered in passing in the above two examples.

\subsection{Example for Section~\ref{sec3}} \label{subsec7.2}
\mbox{} \\[-9mm]

We compute, using Theorem~3.1 and the notes thereafter, the Zernike expansion of $Z_0^0\,\ast\ast_{{\rm corr}}\,Z_0^0$. It is easy to show by elementary means, $Z_0^0\,\ast\ast_{{\rm corr}}\,Z_0^0$ being the area of the common part of two disks of radius 1 whose centers are at a distance $\rho$ apart, that
\beq \label{e164}
(Z_0^0\,\ast\ast_{{\rm corr}}\,Z_0^0)(\rho)=2\Bigl[{\rm arccos}(\tfrac12\rho)-\tfrac12\rho\,\sqrt{1-(\tfrac12\rho)^2}\,\Bigr]~,~~~~~~ 0\leq\rho\leq2~.
\eq
From Eq.~(\ref{e62}) we have
\beq \label{e165}
(Z_0^0\,\ast\ast_{{\rm corr}}\,Z_0^0)(\rho)=2\pi\,\dil_0^{\infty}\,\dfrac{J_1^2(u)\,J_0(\rho u)}{u}\,du~.
\eq
This integral can be found in \cite{ref37}, 2.12.42, item~31 on p.~232, and this would yield Eq.~(\ref{e164}) when the parentheses would have been placed correctly in this reference (a cross-check with \cite{ref37}, 2.12.42, item~15 on p.~230, that arises when the integral on the right-hand side of Eq.~(\ref{e165}) is differentiated with respect to $\rho$, shows inconsistency of \cite{ref37} in this matter).

We have $m''=m-m'=0$ and $n''$ is even at the right-hand side of Eq.~(\ref{e63}), and this yields
\beq \label{e166}
(Z_0^0\,\ast\ast_{{\rm corr}}\,Z_0^0)(\rho)=\dfrac{\pi}{4}\,\dsum_{n''\,{\rm even},{\geq}0}\,(n''+1)\, \Gamma_{00n''}^{000}\,Z_{n''}^0(\tfrac12\rho)~,~~~~~~0\leq\rho\leq2~,
\eq
where
\begin{eqnarray} \label{e167}
\Gamma_{00n''}^{000} & = & 8({-}1)^{-\frac12 n''}\,\dil_0^{\infty}\,J_1(u)\,J_1(u)\,J_{n''+1}(2u)\, \dfrac{du}{u^2}~= \nonumber \\[3.5mm]
& = & 2({-}1)^{\frac12 n''}\,[Q_{00}^{n''+1}+2Q_{02}^{n''+1}+Q_{22}^{n''+1}]~,
\end{eqnarray}
with
\begin{eqnarray} 
& \mbox{} & Q_{00}^{n''+1}=\tfrac12\,(P_{\frac12 n''}^{(0,0)}(0))^2~,\hspace*{4cm}n''=0,2,...~, \\[3.5mm]
& & Q_{02}^{n''+1}=\tfrac18\,P_{\frac12 n''-1}^{(0,2)}(0)\,P_{\frac12 n''-1}^{(2,0)}(0)~,\hspace*{2.4cm}n''=2,4,...~, \\[3.5mm]
& & Q_{22}^{n''+1}=\dfrac{(\frac12 n''+2)!\,(\frac12 n''-2)!}{((\frac12 n'')!)^2}\,\tfrac{1}{32}\, P_{\frac12(n''-2)}^{(2,2)}~,~~~~n''=4,6,...~,
\end{eqnarray}
while $Q_{02}^1$ and $Q_{22}^1$, $Q_{22}^3$ vanish. Thus this yields the Zernike $^0$-expansion of the function
\beq \label{e171}
2({\rm arccos}\,\tau-\tau\,\sqrt{1-\tau^2}\,)=\dsum_{n''\,{\rm even},{\geq}0}\,C_{n''}^0\,Z_{n''}^0(\tau=\frac12\rho)~,~~~~~~0\leq\tau\leq1~.
\eq
The $C_{n''}^0$ are given in integral form as
\beq \label{e172}
C_{n''}^0=4(n''+1)\,\dil_0^1\,({\rm arccos}\,\tau-\tau\,\sqrt{1-\tau^2})\,R_{n''}^0(\tau)\,\tau\,d\tau~;
\eq
the evaluation of the integrals in Eq.~(\ref{e173}) becomes cumbersome, already for low values of $n''=0,2,...\,$. We compute from Eqs.~(\ref{e167})--(171) and $C_{n''}^0=\frac{\pi}{4}\,(n''+1)\,\Gamma_{00n''}^{000}$
\beq \label{e173}
C_0^0=\dfrac{\pi}{4}\,,~~C_2^0={-}\,\dfrac{3\pi}{8}\,,~~C_4^0=\dfrac{5\pi}{32}\,,...~.
\eq
Unfortunately, there does not seem to exist a closed formula for the values of $P_k^{(\gamma,\delta)}(0)$ as required in Eqs.~(169)--(171), except for the case $\gamma=\delta=0$, see \cite{ref14}, Table 22.4.1 on p.~777. Furthermore, the $C_{n''}^0$ decay only slowly because of non-smooth behaviour of $(Z_0^0\,\ast\ast_{{\rm corr}}\,Z_0^0)(\nu,\mu)$ around $(\nu,\mu)=(0,0)$ and, to a lesser extent, around $\nu^2+\mu^2=1$.

\subsection{Examples for Section~\ref{sec4}} \label{subsec7.3}
\mbox{} \\[-9mm]

We have computed, using Theorem 4.1, the Fourier coefficients $a_k$ in the cosine representation
\beq \label{e174}
R_n^m(\cos x)=\dsum_{j=0}^{\lfloor n/2\rfloor}\,a_{n-2j}\cos(n-2j)\,x
\eq
for various cases of integer $n$, $m$ with $n$, $m$ non-negative and $n-m$ even and non-negative. The results are collected in Table~I.

\begin{figure}[h]
 \begin{center}
 \subfigure[]{ \includegraphics[width = 0.48\linewidth]{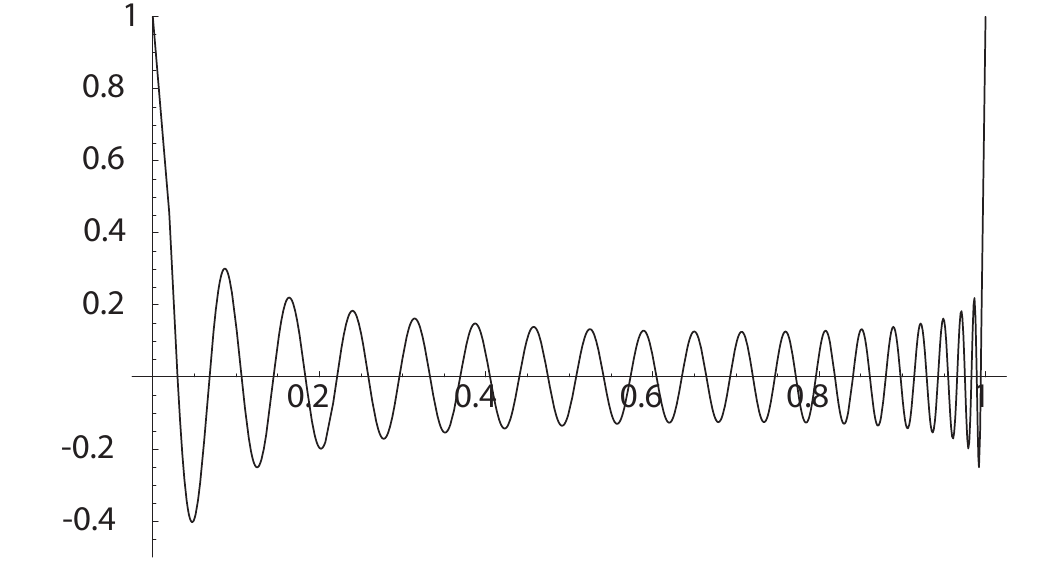}}\hspace{0.1cm}
 \subfigure[]{ \includegraphics[width = 0.48\linewidth]{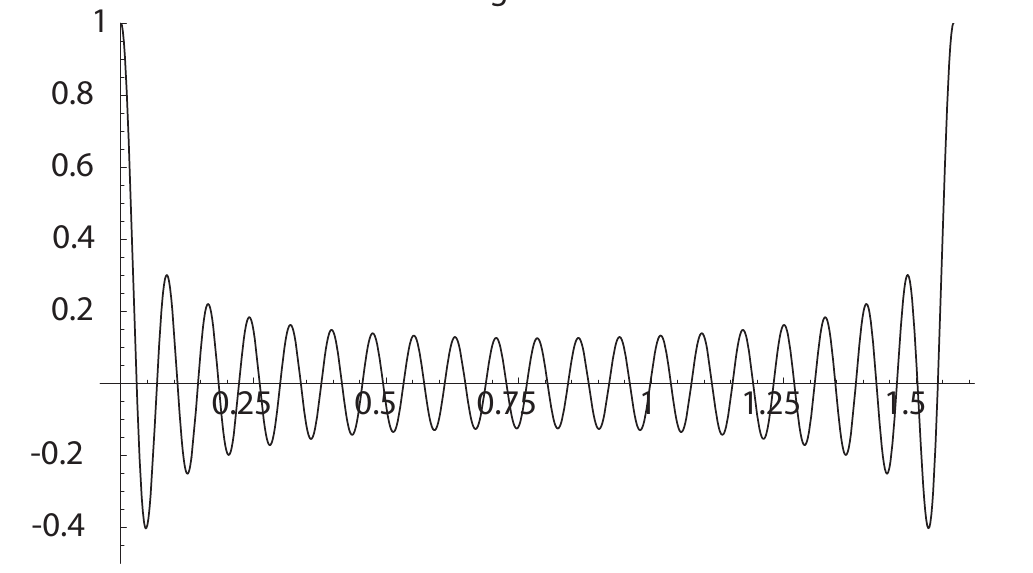}}
 \end{center}
 \caption{Plot of (a) $R_{80}^0(\rho)$, $0\leq\rho\leq1$, and (b) $R_{80}^0(\cos x)$, $0\leq x\leq\pi/2$. The sampling rate used to display (a) is not high enough to adequately represent the last peak but one just before $\rho=1$.}
\end{figure}

\begin{center}
TABLE I. \\
$R_n^m$ in polynomial and cosine representation with $\rho=\cos x$ and $c_k=\cos kx$
\end{center}
$R_0^0~:~~1=c_0$ \\[2.5mm]
$R_2^0~:~~2\rho^2-1=c_2$ \\[2.5mm]
$R_4^0~:~~6\rho^4-6\rho^2+1=\frac14\,c_0+\frac34\,c_4$ \\[2.5mm]
$R_6^0~:~~20\rho^6-30\rho^4+12\rho^2-1=\frac38\,c_2+\frac58\,c_6$ \\[2.5mm]
$R_8^0~:~~70\rho^8-140\rho^6+90\rho^4-20\rho^2+1=\frac{9}{64}\,c_0+\frac{5}{16}\,c_4+\frac{35}{64}\,c_8$ \\[2.5mm]
$R_1^1~:~~\rho=c_1$ \\[2.5mm]
$R_3^1~:~~3\rho^3-2\rho=\frac14\,c_1+\frac34\,c_3$ \\[2.5mm]
$R_5^1~:~~10\rho^5-12\rho^3+3\rho=\frac14\,c_1+\frac18\,c_3+\frac58\,c_5$ \\[2.5mm]
$R_7^1~:~~35\rho^7-60\rho^5+30\rho^3-4\rho=\frac{9}{64}\,c_1+\frac{15}{64}\,c_3+\frac{5}{64}\,c_5 +\frac{35}{64}\,c_7$ \\[2.5mm]
$R_2^2~:~~\rho^2=\frac12\,c_0+\frac12\,c_2$ \\[2.5mm]
$R_4^2~:~~4\rho^4-3\rho^2=\frac12\,c_2+\frac12\,c_4$ \\[2.5mm]
$R_6^2~:~~15\rho^6-20\rho^4+6\rho^2=\frac{3}{16}\,c_0+\frac{1}{32}\,c_2+\frac{5}{16}\,c_4+\frac{15}{32}\,c_6$ \\[2.5mm]
$R_8^2~:~~56\rho^8-105\rho^6+60\rho^4-10\rho^2=\frac{9}{32}\,c_2+\frac{1}{16}\,c_4+\frac{7}{32}\,c_6 +\frac{7}{16}\,c_8$ \\[2.5mm]
$R_3^3~:~~\rho^3=\frac34\,c_1+\frac14\,c_3$ \\[2.5mm]
$R_5^3~:~~5\rho^5-4\rho^3=\frac18\,c_1+\frac{9}{16}\,c_3+\frac{5}{16}\,c_5$ \\[2.5mm]
$R_7^3~:~~21\rho^7-30\rho^5+10\rho^3=\frac{15}{64}\,c_1+\frac{1}{64}\,c_3+\frac{27}{64}\,c_5+\frac{21}{64}\,c_7$ \\[2.5mm]
$R_4^4~:~~\rho^4=\frac38\,c_0+\frac12\,c_2+\frac18\,c_4$ \\[2.5mm]
$R_6^4~:~~6\rho^6-5\rho^4=\frac{5}{16}\,c_2+\frac12\,c_4+\frac{3}{16}\,c_6$ \\[2.5mm]
$R_8^4~:~~28\rho^8-42\rho^6+15\rho^4=\frac{5}{32}\,c_0+\frac{1}{16}\,c_2+\frac18\,c_4+\frac{7}{16}\,c_6+\frac{7}{32} \,c_8$ \\[2.5mm]
$R_5^5~:~~\rho^5=\frac58\,c_1+\frac{5}{16}\,c_3+\frac{1}{16}\,c_5$ \\[3mm]
$R_7^5~:~~7\rho^7-6\rho^5=\frac{5}{64}\,c_1+\frac{27}{64}\,c_3+\frac{25}{64}\,c_5+\frac{7}{64}\,c_7$ \\[3mm]
$R_6^6~:~~\rho^6=\frac{5}{16}\,c_0+\frac{15}{32}\,c_2+\frac{3}{16}\,c_4+\frac{1}{32}\,c_6$ \\[2.5mm]
$R_8^6~:~~8\rho^8-7\rho^6=\frac{7}{32}\,c_2+\frac{7}{16}\,c_4+\frac{9}{32}\,c_6+\frac{1}{16}\,c_8$ \\[3mm]
$R_7^7~:~~\rho^7=\frac{35}{64}\,c_1+\frac{21}{64}\,c_3+\frac{7}{64}\,c_5+\frac{1}{64}\,c_7$ \\[2.5mm]
$R_8^8~:~~\rho^8=\frac{35}{128}\,c_0+\frac{7}{16}\,c_2+\frac{7}{32}\,c_4+\frac{1}{16}\,c_6+\frac{1}{128}\,c_8$ \\ \\
In Fig.~6 we have displayed $R_{80}^0(\rho)$, $0\leq\rho\leq1$, and $R_{80}^0(\cos x)$, $0\leq x\leq\pi/2$ to
illustrate the point that the variation of the radial polynomial is more or less spread out uniformly over the $x$-interval.

\section*{A~~Convergence of the integral in Eq.~(\ref{e7})} \label{appA}
\mbox{} \\[-9mm]

We shall show in this appendix that for non-negative integers $n$ and $m$ with $n-m$ even and non-negative, the integral
\beq \label{e175}
\dlim_{v\pr\infty}\,\dil_0^v\,J_{n+1}(u)\,J_m(\rho u)\,du
\eq
converges to $({-}1)^{\frac{n-m}{2}}\,R_n^m(\rho)$ for $0\leq\rho<1$ and to 0 for $\rho>1$, and that it does so boundedly in $\rho\geq0$ and uniformly in $\rho$ outside $(1-\eps,1+\eps)$ for any $\eps>0$. We have from \cite{ref14}, 9.2.1 on p.~364
\beq \label{e176}
J_k(u)=\sqrt{\dfrac{2}{\pi u}}\cos(u-\frac12\,k\pi-\frac14\,\pi)+O(u^{-3/2})~,~~~~~~u\pr\infty~,
\eq
and $J_k(u)$ is smooth and bounded on $u\geq0$. Therefore, to show bounded and uniform convergence of the integral in Eq.~(\ref{e175}) on the appropriate sets of $\rho$, it is sufficient to establish this for the integral
\beq \label{e177}
\dlim_{v\pr\infty}\,\dfrac{2}{\pi}\,\dil_0^v\,\dfrac{1}{u}\cos(u-\frac12(n+1)\,\pi-\frac14\pi)\cos(\rho u-\frac12 m\pi-\frac14\pi)\,du~.
\eq
Once this has been established, the issue of to what the integral in Eq.~(\ref{e175}) converges is settled by the remark that $Z_n^m$ and $2\pi i^n\exp(im\varp)\,J_{n+1}(2\pi r)/2\pi r$, see Eq.~(\ref{e5}), are $2D$ Fourier pairs so that Fourier inversion of the latter function yields the former in $L^2(\dR)$-sense while the former function is smooth outside the set $\nu^2+\mu^2=1$.

Using elementary trigonometric identities, we have
\begin{eqnarray} \label{e178}
& \mbox{} & \dfrac{2}{\pi}\,\dil_1^v\,\dfrac{1}{u}\cos(u-\frac12 (n+1)\,\pi-\frac14\pi)\cos(\rho u-\frac12 m\pi-\frac14 \pi)\,du~= \nonumber \\[3.5mm]
& & =~\dfrac{({-}1)^p}{\pi}\,\dil_1^v\,\dfrac{\sin(1-\rho)\,u}{u}\,du-\dfrac{({-}1)^q}{\pi}\,\dil_1^v\,\dfrac {\cos(1+\rho)\,u}{u}\,du~= \nonumber \\[3.5mm]
& & =~\dfrac{({-}1)^p}{\pi}\,\dil_{1-\rho}^{(1-\rho)v}\,\dfrac{\sin x}{x}\,dx-\dfrac{({-}1)^q}{\pi}\, \dil_{1+\rho}^{(1+\rho)v}\,\dfrac{\cos x}{x}\,dx~,
\end{eqnarray}
where we have set $p=\frac12\,(n-m)$, $q=\frac12\,(n+m)$. Since both functions
\beq \label{e179}
\dil_0^y\,\dfrac{\sin x}{x}\,dx\,,~~y\geq0~;~~~~~~\dil_1^y\,\dfrac{\cos x}{x}\,dx\,,~~y\geq1~,
\eq
are bounded and have a finite limit as $y\pr\infty$, the convergence of the integral in Eq.~(\ref{e177}) is bounded in $\rho\geq0$ and uniform in any closed set of $\rho$'s not containing 1. The assumption that $n$ and $m$ have same parity is essential: $\il_1^{\infty}\,\frac{1}{u}\cos^2u\,du=\infty$. \\ \\
{\bf Acknowledgements.} \\
The author wishes to express his thanks to Prof.\ Joseph Braat for his continuous interest and encouragement when this research was carried out, to Prof.\ Erik Koelink for calling the reference \cite{ref25} to the author's attention and for bringing the result of \cite{ref25} in appropriate context, and to Prof.\ Ronald Aarts with whom the research on transient responses in Section~\ref{sec5} is carried out in the acoustical context. This research was supported by Agentschap NL through the KWR-project Metrology.

\end{document}